\documentclass[%
10pt,A4paper,twocolumn,aps,prl, reprint, superscriptaddress,longbibliography, nofootinbib,
 amsmath,amssymb,
]{revtex4-2}
\usepackage{lipsum,etoolbox}
\usepackage{mathtools}
\usepackage{graphicx}
\usepackage{dcolumn}
\usepackage{bm}
\usepackage{xcolor}
\usepackage{siunitx}
\usepackage{booktabs} 
\usepackage{array}
\usepackage{calrsfs}
\usepackage{multirow}
\usepackage{makecell}
\usepackage{enumitem}
\usepackage{placeins}
\usepackage{comment}
\usepackage[shortcuts]{extdash}
\usepackage[colorlinks=true, linkcolor=blue, citecolor=black, urlcolor=blue, pdfauthor={Yongxin Song}, pdftitle={bqc_manuscript}]{hyperref}

\DeclareMathAlphabet{\mathcal}{OMS}{cmsy}{m}{n}
\setlist[itemize]{leftmargin=*}

\begin{document}
\preprint{APS/123-QED}

\title{Blind Quantum Computation on a Modular Superconducting Processor}

\author{Yongxin Song}
\email{yongxin.song@phys.ethz.ch}
\author{Johannes Knörzer}
\author{Kieran Dalton}
\author{Andreas Wallraff}
\author{Jean-Claude Besse}
\affiliation{Department of Physics, ETH Zurich, CH-8093 Zurich, Switzerland}
\affiliation{Quantum Center, ETH Zurich, CH-8093 Zurich, Switzerland}
\affiliation{ETH Zurich - PSI Quantum Computing Hub, Paul Scherrer Institute, CH-5232 Villigen, Switzerland}

\begin{abstract}
Current cloud-based quantum processors offer access to advanced hardware hosted on a remote server, but do not guarantee data or algorithm privacy. 
Blind quantum computation provides information-theoretic privacy by enabling a client to execute an algorithm without disclosing information about either the task or the final result.
Here, we execute a measurement-based blind quantum computation protocol on a superconducting processor comprising two flip-chip-bonded modules, one acting as a server and the other as a client. 
The server generates a two-dimensional cluster state and forwards it to the client.
Using this resource, the client implements a universal gate set with only adaptive single-qubit rotations and measurements.
To illustrate this approach, we execute a three-qubit instance of the Deutsch-Jozsa algorithm. 
We analyze the server's quantum state after each rotation of a measurement-based single-qubit gate to verify that negligible information about the computation is revealed to the server, consistent with the one-way flow of information that guarantees blindness.
This proof-of-principle demonstration establishes key elements of blind quantum computation in superconducting-circuit architectures, indicating that intermediate-scale implementations of blind protocols may become feasible with realistic near-term improvements in gate fidelities.
\end{abstract}

\date{\today}

\maketitle
\setcounter{secnumdepth}{3}

Quantum computers may execute algorithms that outperform all known classical algorithms for specific tasks, such as integer factorization~\cite{Gidney2025b}, Hamiltonian simulation~\cite{Low2019a}, and topological data analysis~\cite{Berry2024a}.
Although many challenges remain to realize practically useful quantum computation beyond classical capabilities, remarkable progress has been made in scaling quantum systems to hundreds and thousands of qubits~\cite{Ransford2025, Chiu2025a, Acharya2025}.
In addition, several experiments have demonstrated increasingly complex quantum simulation and error correction protocols that highlight the growing capabilities of modern quantum platforms~\cite{Acharya2025, Bluvstein2023, Abanin2025}.
At the same time, cloud-based access to quantum computing hardware has been available to early adopters for a decade~\cite{Devitt2016, Blinov2021, Wurtz2023}.

Using the cloud, users can upload quantum circuits to remote servers and receive measurement results in minutes.
Yet in scenarios where data privacy is a concern, the user may not want to reveal their input data, the output, or the computation itself.
This motivated the study of blind quantum computation, which enables a client to delegate a computation while keeping both the computation and its result hidden from the provider~\cite{Childs2005b,Arrighi2006}.

Measurement-based quantum computation~\cite{Raussendorf2001} provides a natural framework for blind quantum computation~\cite{Fitzsimons2017}.
One way to achieve blindness is for the client to prepare randomly rotated single-qubit states and send them to the server, which performs entangling operations and measurements and reports outcomes classically~\cite{Broadbent2009}.
Here we instead consider the setting in which the server prepares an entangled state and sends it to the client, who performs adaptive measurements to implement universal quantum computation~\cite{Morimae2013}.
This protocol is by construction guaranteed to be blind, as all communication proceeds from server to client.

An early blind quantum computation experiment prepared four-qubit cluster states of optical photons, demonstrating blind execution of small instances of Deutsch's and Grover's algorithms~\cite{Barz2012}.
Further blind quantum computation demonstrations capitalized on the availability of photonic quantum links~\cite{Barz2013,Fisher2014,Greganti2016,Huang2017}.
Although matter-based qubits have long been considered as a promising platform for quantum computation because they enable high-fidelity storage and manipulation of quantum information, experimental demonstrations of blind quantum protocols on matter-based quantum processors have only recently appeared, including implementations with trapped ions~\cite{Drmota2024} and silicon-vacancy centers~\cite{Wei2025}.

Superconducting circuits provide an alternative scalable platform, where first steps toward measurement-based quantum computation have been demonstrated on cloud-based monolithic devices, with logical operations realized through classical post-processing~\cite{Pathumsoot2020, Jiang2026} or by employing additional delayed-choice circuits~\cite{Yang2022b}.
Implementing deterministic measurement-based quantum computation with real-time feedforward control will enable a more efficient and scalable route to extending such computations to large-scale quantum devices.

\begin{figure}[b]\setlength{\hfuzz}{1.1\columnwidth}
\begin{minipage}{\textwidth}
\includegraphics[width=\textwidth]{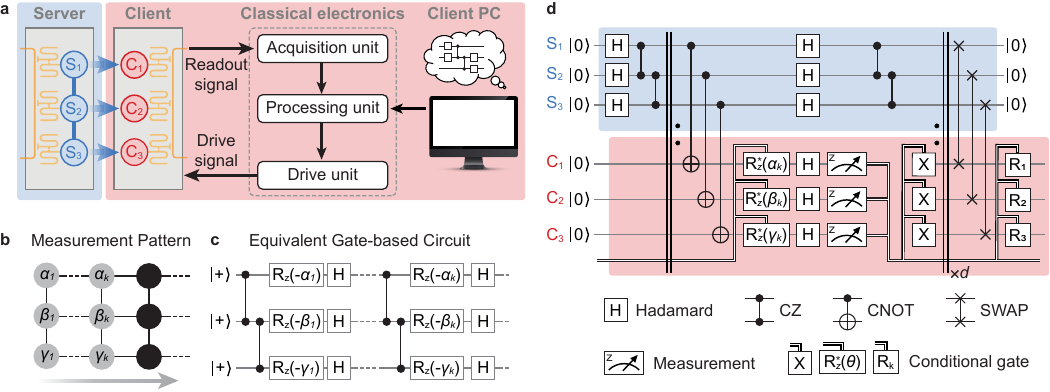}
\caption{\label{fig:fig1_principle}{\bf Blind quantum computation with a modular superconducting quantum device.} 
{\bf (a)} Principle of the experiment.
The experimental device consists of a server (blue) and a client (red) quantum module, each hosting three transmon qubits (circles) having dedicated control (not shown in the illustration) and readout (yellow) circuits.
The server generates the resource cluster state and transfers it to the client. 
In a feedforward experiment, the readout signal is acquired and processed by the instrument in real-time which generates the conditional drive pulses.
Information about the target computation sequence is only acquired by the classical instruments on the client side and not revealed to the server.
{\bf (b)} An example measurement pattern. Gray circles denote the measured cluster state nodes and black circles denote the nodes to be measured. 
Angles $\theta\in\{\alpha_k,\beta_k,\gamma_k\}$ represent the measurement in basis $\{(|0\rangle \pm e^{i\theta} |1\rangle)/\sqrt{2}\}$. 
The arrow shows the sequence of the measurements. 
{\bf (c)} Equivalent circuit diagram of the measurement pattern. 
{\bf (d)} Gate sequence realizing a generic measurement-based protocol. 
The circuit between the repeat sign is executed multiple times to successively generate layers of cluster states at the server side and measure those at the client side.
} 
\end{minipage}
\end{figure}

Here, we realize a deterministic blind quantum computation protocol on a modular superconducting quantum processor that comprises a server and a client module.
The server generates a resource cluster state and transmits it to the client which carries out arbitrary quantum computations through measurements and adaptive single-qubit rotations only.
Within this framework, we realize measurement-based single- and two-qubit gates, as well as a three-qubit instance of the Deutsch-Jozsa algorithm.
Under the assumption that no information is leaked to the server via unintended pathways, this protocol is blind, such that the computation and its results are known only to the client.
We quantify the information leakage by characterizing the server’s quantum state after each rotation of a measurement-based single-qubit gate, revealing negligible information about the client’s state.
The experiment constitutes a proof-of-principle demonstration of blind quantum computation in a modular superconducting architecture, where the client and the server are realized as separate modules within a single device.
Our analysis suggests that intermediate-scale blind quantum computation could become feasible with realistic near-term improvements in gate performance.

\section*{Modular setup and gate sequence}\label{sec:modular_architecture_and_measurement_pattern}

The server and client modules of our superconducting quantum device each host three flux-tunable transmon qubits~\cite{Koch2007}. 
The modules are flip-chip bonded to a common carrier chip that integrates the control and readout circuits in a device similar to the one presented in Ref.~\cite{Dalton2025}. 
Each qubit has an individual drive and flux line, enabling independent state manipulation and frequency tuning. 
A dedicated frequency-multiplexed readout line is used on each module to dispersively read out each qubit.
Static qubit–qubit couplers are implemented both within and between the modules to enable flux-activated two-qubit gates~\cite{Norris2026}. 
We use a real-time feedforward control system at the client which processes individual qubit readout signals and forwards the result to a processing unit that instructs the drive electronics to generate the feedforward pulses according to the user-programmed logic, see Fig.~\ref{fig:fig1_principle}a.
We discuss the experimental setup in more detail in Appendix~\ref{app:sample_and_experimental_setup}.

In our experiment, we implement measurement-based quantum computation by generating cluster states using the server module and measuring them sequentially on the client module.
A given quantum computation is specified by a measurement pattern, defined by the topology of the cluster state and the chosen measurement bases.
The measurement bases must be updated in real time according to the outcomes of previous measurements to obtain a deterministic result~\cite{Raussendorf2003}.
Although the size of the resource state is fixed at the server, neither the computation nor its outcome is revealed.
The separate assignments of server-side resource-state preparation and client-side adaptive measurements form the basis of blind quantum computation in the protocol that we consider.

We realize two-dimensional cluster states of sizes $w\times d$, with a width $w\in\{1,2,3\}$ and a depth $d\in\mathbb{Z}^+$.
Given the specific resource state, we specify each step of the computation executed at the client by a set of three projection angles, $\{\alpha_k, \beta_k, \gamma_k\}$, see Fig.~\ref{fig:fig1_principle}b.
Each angle $\theta \in \{\alpha_k, \beta_k, \gamma_k\}$ defines an orthonormal measurement basis corresponding to an axis in the equatorial plane of the Bloch sphere, given by $B(\theta) \equiv \{|0\rangle + e^{i\theta}|1\rangle, |0\rangle -$
\linebreak
\newpage
\noindent 
$e^{i\theta}|1\rangle \}/\sqrt{2}$.
Measurement-based quantum computation can also be performed with subsets of the qubits in the described cluster states by leaving the unused qubits in the ground state.

Each measurement-based computation can be mapped onto an equivalent gate-based circuit. 
The measurement-based quantum computation proceeds horizontally along the time axis, where each row of the graph corresponds to a qubit register
within the equivalent circuit description, see Fig.~\ref{fig:fig1_principle}c. 
The vertical edges between adjacent vertices denote controlled-$Z$ (C$Z$) gates, and translate directly into C$Z$ operations between qubits in the gate-based circuit~\cite{Nielsen2004}. 
Measuring a client qubit in a basis, $B(\theta)$, is equivalent to applying the single-qubit unitary $X^s H R_z(-\theta)$ to the state of the qubit in the gate-based circuit, where $s \in \{0,1\}$ denotes the measurement outcome~\cite{Nielsen2006}.
The equivalent gate-based circuit when all measurement byproduct operators $X^s$ are accounted for therefore comprises nested layers of C$Z$, single qubit z-rotations and H gates, see Fig.~\ref{fig:fig1_principle}c.

We implement the measurement pattern in Fig.~\ref{fig:fig1_principle}b with the modular device, see Fig.~\ref{fig:fig1_principle}d. 
First, we create a $w \times 1$ cluster state at the server module. 
Then, we transfer the entangled state to the client by applying CNOT gates between the server and client modules, while preserving the inter-module entanglement required for the subsequent round of state transfer.
Upon receiving the resource state, we apply adaptive basis rotation pulses $HR^*_z(\theta)$ and measure the qubits in the computational basis at the client.
X gates are applied conditioned on the measurement result to reset the client qubits to the ground state, preparing for the next state transfer. 
In the meantime, we generate another layer of the cluster state using the qubits at the server. 
The entangling operations, state transfer, and mid-circuit measurements are executed $d$ times until the computation is complete. 
This iterative protocol enables measurement-based quantum computation with cluster state sizes $n=w\times d$ beyond the physical qubit number on the experimental device, while ultimately being limited by gate and decoherence errors in the sequence.
Finally, a circuit swaps the quantum state of the server register with that of the client, and the client applies the final feedforward pulses to generate the target output state deterministically. 

\section*{Resource state characterization}
\begin{figure}[t!]
    \includegraphics{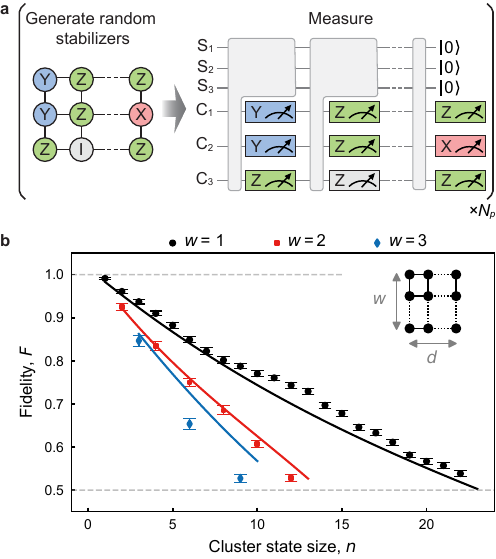}
    \caption{\label{fig:fig2_resource_state_characterization}{\bf Fidelity estimation of the resource cluster state.} 
    {\bf(a)} Experimental sequence. 
    Left: Randomly generating a stabilizer. Each circle represent a qubit in the cluster state. The stabilizer is expressed as the product of single-qubit $X$ (red), $Y$ (blue), $Z$ (green), and identity (gray) operators acting on each qubit.
    Right: Measurement of the stabilizer by updating the measurement bases of the mid-circuit and final measurement steps. 
    The generation and measurement is repeated $N_p$ times for estimating the fidelity of the cluster state.
    {\bf(b)} Estimated fidelities of cluster states with width $w$, depth $d$, and size $n=w\times d$, see inset for the illustration of a generic cluster state.  
    Measured cluster state fidelity with width $w=1,2,3$ are shown with black circles, red squares, and blue diamonds respectively. 
    Error bars are obtained from bootstrapping analysis. 
    Curves of the same color show the fidelity determined using circuit-level simulation. Horizontal dashed lines at $F=1.0$ and $0.5$ mark the fidelity of perfect resource states and the threshold for genuine multipartite entanglement.
    }
\end{figure}

Since the fidelity of the cluster state directly impacts the fidelity of the measurement-based quantum computation, we characterize the resource state using an efficient fidelity estimation method~\cite{Flammia2011} based on a random selection of stabilizers which can be evaluated via Pauli measurements.
This method is particularly well-suited for stabilizer states, including cluster states, enabling unbiased fidelity estimation of arbitrarily large cluster states with constant sampling overhead, provided that measurement errors are sufficiently small.

We prepare and characterize cluster states of varying size $n$. 
For $n \leq 8$, we repeatedly loop over all stabilizers of the cluster state and sample 256 operators.
For $n>9$, we randomly and uniformly sample $N_\mathrm{p}=400$ stabilizers, see Fig.~\ref{fig:fig2_resource_state_characterization}a and details in Methods~\ref{method:fidelity_estimation_of_the_resource_cluster_state}. 
Each stabilizer is measured 8192 times, and the stabilizer expectation value is calculated after applying readout error mitigation, see Appendix~\ref{app:readout_error_correction}.
First, we generate $1\times d$ cluster states with $d\in\{1,\dots,22\}$. 
We utilize one pair of server and client qubits and repeat the iterative generation and measurement of the cluster state until the target depth, $d$, is reached. 
We average the measured stabilizer expectation values to estimate the fidelity.
We find that all created cluster states have fidelities greater than 0.5, implying genuine multipartite entanglement~\cite{Toth2005}, see Fig.~\ref{fig:fig2_resource_state_characterization}b (black circles).
We extract a fidelity $F=0.538(8)$ for the 1D 22-qubit cluster state.
Then, we proceed to $2\times d$ (red squares) and $3\times d$ (blue diamonds) cluster states and report $F=0.528(7)$ for a $n=2\times 6$ state and $F=0.53(1)$ for a $n=3\times 3$ state.
The measured fidelity quantitatively agrees with circuit-level simulations (lines) including quantum gate errors, idling errors and readout errors, see Appendix~\ref{app:numerical_simulation_of_mbqc_circuit}.

Generation of cluster states with increasing $w$ is more susceptible to error because most of the two-qubit gates must be performed sequentially. As the sequence length increases, the qubits become more strongly affected by decoherence and residual $ZZ$ interactions during the idle periods.
We present the experimental sequence and the time budget in Appendix~\ref{app:gate_sequence_and_time_budget}.

\section*{Deterministic single- and two-qubit gates}\label{sec:measurement_based_gates}

\begin{figure}[t!]
    \includegraphics{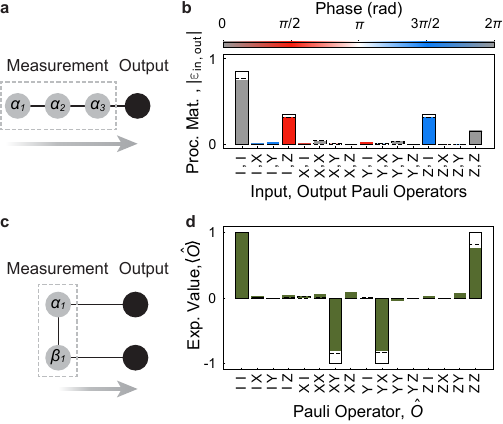}
    \caption{\label{fig:fig3_measurement_based_gates}{\bf Deterministic implementation of measurement-based single- and two-qubit gates.} {\bf (a)} Measurement pattern used to realize arbitrary single-qubit unitaries. 
    The cluster state nodes (gray disks) are measured in the sequence indicated by the arrow, resulting in the output state (black).
    {\bf (b)} Experimental process matrix $|\varepsilon_\mathrm{in, out}|$ of a deterministic measurement-based T gate (filled bars), realized by setting $\alpha_1=-\pi/4, \alpha_2=\alpha_3=-\pi/2$ in (a). 
    The ideal and simulated processes are shown with solid and dashed wire frames.
    {\bf (c)} Measurement-based generation of the output state  $|\psi\rangle = (|00\rangle - i |11\rangle)/\sqrt{2}$. Here, $\alpha_1 = \beta_1 = -\pi /2$. 
    {\bf (d)} Measured Pauli operator expectation values of $|\psi\rangle$ prepared with the measurement-based circuit.
    }
\end{figure}

We demonstrate measurement-based single- and two-qubit gates as the essential primitives for realizing measurement-based quantum computation.
First, we realize arbitrary single-qubit rotations whose unitaries can be expressed as a sequence of rotations around the $x$- and $z$-axes, parameterized by the three Euler angles $\alpha_1$, $\alpha_2$, and $\alpha_3$.
To accomplish a rotation specified by these three angles, we measure the first three nodes of a four-qubit linear cluster state in bases $B(-\alpha_1),~B(-\alpha_2)$ and $B(-\alpha_3)$, respectively, see Fig.~\ref{fig:fig3_measurement_based_gates}a, and account for the byproduct operators by adapting the subsequent measurement bases, see Methods~\ref{method:compenstaion_measurement_byproduct}.
The output state can be expressed as $|\psi_\mathrm{out}\rangle = H R_z(\alpha_3) R_x(\alpha_2) R_z(\alpha_1)|+\rangle$.
As a building block for universal quantum computation, we implement a deterministic measurement-based T-gate by choosing $\alpha_1 = -\pi / 4, \alpha_2 = \alpha_3 = -\pi / 2$. 
The experimental sequence requires five C$Z$ gates, three mid-circuit measurements, and is executed in a time of $\SI{3.6}{\mu s}$. 
We determine the T-gate process fidelity $F_\mathrm{proc} = \mathrm{Tr}(\varepsilon\varepsilon_\mathrm{ideal}) = 0.902$ using quantum process tomography~\cite{Chuang1997} with readout error mitigation, see Fig.~\ref{fig:fig3_measurement_based_gates}b. 
Here, $\varepsilon_\mathrm{ideal}$ and $\varepsilon$ are the ideal and measured process matrices.
We also characterized measurement-based arbitrary single-qubit unitaries by parametrically sweeping the rotation angles $\alpha_1$ and $\alpha_2$, see Appendix~\ref{app:arbitrary_single_qubit_gate}.

We realize a measurement-based two-qubit gate using a two-dimensional cluster state, such as the one depicted in Fig.~\ref{fig:fig3_measurement_based_gates}c.
By measuring the first column of qubits in the $B(-\alpha_1)$ and $B(-\beta_1)$ bases and accounting for the byproduct operator, the measurement pattern yields the output state $|\psi_\mathrm{out}\rangle = [H^{(1)} R_z^{(1)}(\alpha_1)] \otimes [H^{(2)} R_z^{(2)}(\beta_1)] \, \mathrm{C}Z \, |++\rangle \, $,
where superscripts $(1)$ and $(2)$ denote the single-qubit operations applied to the output qubit in the first and second row, respectively.
We choose $\alpha_1 = \beta_1 = -\pi / 2$, which ideally yields a maximally entangled output state $|\psi\rangle = (|00\rangle - i
|11\rangle) / \sqrt{2}$. 
We perform quantum state tomography~\cite{James2001} on the output state and obtain a fidelity of $F=0.849$, see the measured Pauli operator expectation values in Fig.~\ref{fig:fig3_measurement_based_gates}d. 
We also perform quantum process tomography on the two-qubit operation and extract a process fidelity $F_\mathrm{proc} = 0.845$, see Appendix~\ref{app:quantum_process_tomography}. 
Our circuit-level simulation shows that decoherence during idling is the dominant error source in our measurement-based gates, with additional contributions from readout, C$Z$, and single-qubit gate errors, see Appendix~\ref{app:numerical_simulation_of_mbqc_circuit}.
The demonstrated single- and two-qubit gates establish a universal gate set for measurement-based quantum computation.

\section*{Deutsch-Jozsa algorithm} \label{sec:dj_algorithm}

To illustrate the execution of a quantum algorithm on our modular device and within the measurement-based quantum computation framework, we implement a small instance of the Deutsch-Jozsa algorithm~\cite{Vallone2010, Wei2025}.
It determines whether a Boolean function $f: \{0,1\}^{m}\rightarrow \{0,1\}$ is constant (returns the same output for all inputs) or balanced (returns an equal number of zeros and ones).

\begin{figure}[t!]
    \includegraphics{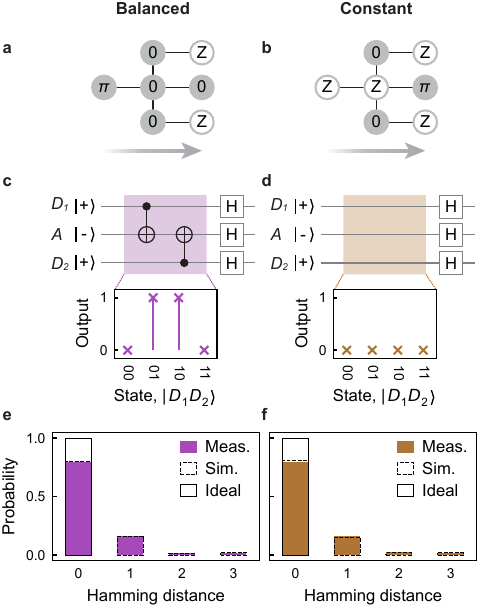}
    \caption{\label{fig:fig4_three_qubit_protocol}{\bf A measurement-based Deutsch-Jozsa algorithm.} 
    Measurement pattern of a Deutsch-Jozsa algorithm with {\bf (a)} balanced and {\bf (b)} constant oracles. 
    Filled circles with notations 0 and $\pi$ mark the qubits measured in the $B(0)$ and $B(\pi)$ bases.
    Empty circles with notation $Z$ mark the qubits measured in the computational basis.
    The arrow indicates the sequence in which the measurements are processed.
    Equivalent circuit of the \textbf{}Deutsch-Jozsa algorithm with {\bf (c)} balanced and {\bf (d)} constant oracles. 
    The ideal output states of the oracles are presented below the circuits. 
    Measured output state probabilities (filled bars) with {\bf (e)} balanced and {\bf (f)} constant oracles, plotted with the Hamming distance with respect to the ideal output. 
    Dashed wire frames show the simulation result.
    Solid wire frames show the ideal output.
    }
\end{figure}

We realize the Deutsch-Jozsa algorithm on $m=2$ input bits using the standard circuit formulation with $m$ data qubits and one auxiliary qubit, resulting in a total of three qubits.
For this case, there exist two constant and six balanced oracles, representing the possible functions $f$~\cite{Tame2010}.
As examples, we implement one of the constant oracles and one balanced oracle.
For the balanced oracle, we choose a two-bit parity check function. The constant oracle that we realize always outputs $1$.
As a resource state we employ the seven-qubit cluster states shown in Fig.~\ref{fig:fig4_three_qubit_protocol}a,~b, of which the qubits are measured in the sequence indicated by the arrow in either of the bases $B(0),~B(\pi)$, or in the computational basis.
The equivalent gate-based quantum circuits are shown in Fig.~\ref{fig:fig4_three_qubit_protocol}c,~d. 
Each row of the cluster state is translated to a qubit register in the circuit picture, denoted as data qubits $D_1, D_2$ and auxiliary qubit $A$.

Ideally, the Deutsch-Jozsa algorithm applied to the balanced oracle should deterministically result in the output bitstring $|D_1 A D_2\rangle = |111\rangle$. With 8,192 executions, we measure the correct bitstring with a probability of 0.797, and results with Hamming distances 1, 2, and 3 from the correct bitstring with probabilities of 0.163, 0.020 and 0.020, respectively, see Fig.~\ref{fig:fig4_three_qubit_protocol}e.
Most of the erroneous output states display a one-bit error, which we attribute to the sequence being dominated by uncorrelated single-qubit errors.
The algorithm with the constant oracle ideally yields the bitstring $|D_1 A D_2\rangle = |010\rangle$.
We observe bitstrings with Hamming distances of 0 to 3 from the correct result occurring with probabilities 0.798, 0.159, 0.023, and 0.020, respectively, showing a distribution similar to that of the balanced oracle, see Fig.~\ref{fig:fig4_three_qubit_protocol}f.

For comparison, if blindness is not required, a gate-based implementation of the Deutsch-Jozsa algorithm on the server can realize the same computation with a shorter sequence of quantum operations~\cite{DiCarlo2009}.
Based on the average operation fidelities of the experimental device, we estimate a success probability of 0.973 for $m=2$ input bits, see discussions in Appendix~\ref{app:performance_and_overhead_analysis_mbqc}.

\section*{Blind quantum computations}\label{sec:blindness_of_the_computing}

A key feature of the demonstrated measurement-based quantum computation framework is that it enables a client to perform quantum computations without revealing the executed algorithm or results to the server.
This requires that the quantum state accessible to the server be independent of the client's computation.
To probe this requirement experimentally, we characterize the quantum state accessible to the server throughout the execution of a quantum operation.
We perform this characterization in the measurement-based implementation of a T-gate, as an example of a nontrivial instance of the adaptive measurement and feed-forward structure of the protocol.

\begin{figure}[t!]
    \centering
    \includegraphics{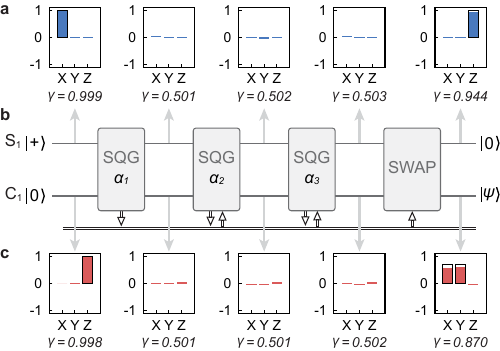}
    \caption{\label{fig:fig5_privacy_characterization}{\bf Verifying the blindness of the protocol.} 
    Measured Pauli operator expectations values of {\bf (a)} the server and {\bf (c)} the client qubit at different stages of a measurement-based T-gate execution, gate sequence in {\bf (b)}. 
    Solid wire frames show the ideal expectation values.
    The purity $\gamma$ of each state is quoted below each tomogram.
    Each of the first three blocks represents generating and measuring one node along the 1D cluster state in the basis $B(\alpha_i), i=1,2,3$. 
    The last block represents the SWAP circuit between the server and the client module.
    }
\end{figure}

At each step of the execution of the T-gate, we locally characterize the server and client qubits with quantum state tomography and extract the expectation values of the corresponding Pauli operators.
With the server qubit deterministically prepared in the $|+\rangle$ state at the beginning of the sequence, we find that it is in a fully mixed state after each measurement-based rotation in the sequence with respective purities of 0.501, 0.502, 0.503, since the measurement eliminates all coherence between $|0\rangle$ and $|1\rangle$.
By the end of the circuit, the server qubit is returned to the $|0\rangle$ state, see Fig.~\ref{fig:fig5_privacy_characterization}a,~b, as designed by the protocol.

Similarly, the client qubit is prepared in the $|0\rangle$ state at the beginning of the sequence, which we find agreement in the measurement.
After each mid-circuit measurement, the client qubit is in the fully mixed state due to the randomness of the projective readout, see Fig~\ref{fig:fig5_privacy_characterization}c.
Finally, after swapping with the server qubit and applying the final feedforward pulse, the client qubit ideally is prepared in the target state $T|+\rangle = (|0\rangle + \mathrm{e}^{i\pi/4}|1\rangle) / \sqrt{2}$.

Furthermore, we quantify the blindness in our implementation of the protocol by measuring the Holevo information~\cite{Holevo1973} available to the server when the client applies several single-qubit rotations around the $y$- and $z$-axes.
This captures information leakage across different client inputs.
By incrementing the y- or z- rotation angles on one of the client qubits in steps of $\pi/8$, we generate $30$ distinct states on the Bloch sphere.
For each of these rotations, we perform quantum state tomography on the corresponding server qubit and reconstruct the resulting ensemble of states, from which we obtain a Holevo information of $2.3(5)\times 10^{-4}$ bits.

The Holevo information provides an upper-bound to the amount of classical information that any measurement on the server qubit could reveal about which state the client prepared.
Since a single qubit can in principle reveal at most $1$ bit of classical information, our experimentally obtained value is orders of magnitude below this limit, see Methods~\ref{method:measuring_info_leakage} for details. 

We adopt the assumption of no unintended information leakage from client to server~\cite{Morimae2013}.
This assumption underlies blind quantum computation protocols, but it is not yet rigorously enforced at the hardware level in current experimental implementations, where residual side channels may in principle still leak information.
While the assumption is consistent with our observation of negligible Holevo information, establishing rigorous privacy guarantees remains an important open challenge.
Future improvements could include increased physical separation and isolation between the client and server modules to suppress unwanted crosstalk and other side channels, together with dedicated leakage characterization. 
At the protocol level, stronger security guarantees could be achieved through extensions such as verifiable delegated-computation schemes~\cite{Fitzsimons2017}.
Under the assumption of no unintended information leakage, blindness in the protocol we have demonstrated here is conceptually ensured through three key aspects.
First, in our experiments the cluster states are prepared on the server module independently of the computation.
Second, when we execute the protocol, at the client we address the qubits on the client module with the control and readout lines unique to that module. 
We generate and acquire the control and readout signals with the trusted instruments only available to the client side of the experiment, which are distinct from the hardware we use on the server side. 
Finally, we process the readout results privately using electronics at the client, which is physically separated form the readout electronics we use at the server side.
Although the server holds the intermediate quantum state of the measurement-based sequence, it cannot reconstruct the target state of the client due to missing mid-circuit measurement results.
The intermediate quantum state obtained by the server is a probabilistic mixture of the computation target states, with and without the byproduct operators, thereby concealing the details of the client’s computation.

While the demonstrated protocol enables blind quantum computation, it introduces a resource overhead compared to performing the same computation publicly on the server.
To quantify this overhead, we compare our measurement-based implementation to equivalent gate-based circuits.
Our analysis accounts for imperfections on both the server and the client sides in preparing, transmitting, processing, and measuring the resource states, and takes into account realistic levels of experimental errors from single-qubit and two-qubit gates, mid-circuit readout, and idling.
We estimate how much the error rates in the measurement-based implementation must be reduced to match the output fidelity of state-of-the-art gate-based performance, and find that this requires lowering those errors by roughly one order of magnitude, with even smaller improvements sufficient for circuits dominated by two-qubit gates.
Further details are provided in Appendix~\ref{app:performance_and_overhead_analysis_mbqc}.
These results indicate that blind quantum computations of intermediate scale may be feasible with realistic near-term experimental improvements.

\section*{Discussion and Outlook}\label{sec:discussion_and_outlook}

In this work, we have experimentally implemented blind quantum computation on a modular superconducting processor. 
We generated cluster states on the server and enabled the client to perform universal blind quantum computation by measuring them using only single-qubit operations. We expect to naturally extend the protocol to clients with fewer physical qubits by implementing couplers from one client qubit to multiple server qubits.

The performance of blind quantum computing protocols could be improved further by incorporating quantum error mitigation and correction into the framework.
For example, fault-tolerant measurement-based quantum computation can be enabled using three-dimensional cluster states~\cite{Raussendorf2007a}, and error-protected quantum interconnects~\cite{Ramette2023} could enable larger distances between clients and servers.

The demonstrated blind quantum computation protocol can be integrated into distributed quantum computing systems.
Servers and clients housed in separate dilution refrigerators could be connected via microwave-frequency quantum links~\cite{Magnard2020, Yam2025} or microwave-to-optical transducers~\cite{Han2021f}, enabling transmission of the resource state through entangled microwave or optical photons~\cite{Schwartz2016, Larsen2019, Ferreira2024, OSullivan2025}.
In this way, cloud-based services could incorporate information-theoretic blindness guarantees.
The prepare-and-measure protocol is platform-independent and has been deployed with other architectures, including neutral atoms, trapped ions, and spin qubits~\cite{Drmota2024, Wei2025}. 
Our work represents a step toward secure user access to universal quantum computation enabled by remote quantum servers.

\section*{Acknowledgments}
The authors thank Pieter-Jan Stas for valuable feedback on the manuscript. The authors acknowledge financial support by the Swiss State Secretariat for Education, Research and Innovation under contract number UeM019-11 and by ETH Zurich.

\section*{Author contribution}
Y.\,S. and \mbox{J.-C.\,B}. planned the experiments. Y.\,S. developed the measurement control sequence and acquired the data. J.\,K, Y.\,S., and K.\,D. contributed to the analysis. K.\,D. contributed to the measurement setups and designed and fabricated the quantum device used for this experiment, A.\,W. and \mbox{J.-C.\,B}. supervised the project. Y.\,S. and J.\,K. wrote the manuscript with inputs from all authors.

\section*{Methods}
\label{app:methods}

\subsection{Fidelity estimation via Pauli measurements}\label{method:fidelity_estimation_of_the_resource_cluster_state}

The fidelity of an experimentally prepared $n$-qubit state $\rho$ with respect to a cluster state $|\psi\rangle$ can be efficiently estimated using Pauli measurements~\cite{Flammia2011, Cao2023a}.
An unbiased estimator of $F=\langle \psi | \rho | \psi \rangle$ is obtained by sampling Pauli operators according to the characteristic function of $|\psi\rangle$ and averaging the corresponding expectation values.

The cluster state is uniquely defined by the generators $\{S_m\}$ of its stabilizer group, with $S_m = X_m\Pi_{l\in N(m)} Z_l$, $m=1,2,\dots, n$ running over all vertices of the cluster state and $N(m)$ the neighborhood of a given vertex $m$.
The stabilizer group $\{P_k\}_{k=1,2,\dots, 2^n}$ is generated by $\{S_m\}$ and contains all $2^n$ distinct products of these commuting generators. 
The stabilizers $P_k$ satisfy $\langle \psi | P_k | \psi \rangle = 1, \, \forall k$, while non-stabilizer Pauli operators have vanishing expectation value.
The fidelity can therefore be estimated by uniformly sampling stabilizers and averaging
$F = \frac{1}{N_\mathrm{p}}\sum_{j=1}^{N_\mathrm{p}} \langle P_j\rangle \,$.

Experimentally, stabilizers are generated by randomly selecting products of the generators.
For $n\leq 8$, we repeatedly sample the full stabilizer group, while for $n>9$ we uniformly sample $400$ stabilizers.
Each stabilizer is measured using single-qubit Pauli measurements, applying a Hadamard gate for $X$ measurements and a $H R_z(\pi/2)$ rotation for $Y$ measurements prior to readout in the computational basis.
The expectation value $\langle P_j \rangle$ of each sampled stabilizer is estimated from $8192$ repetitions using readout error mitigation, see Appendix~\ref{app:readout_error_correction}.
Uncertainties are estimated by bootstrapping the measured expectation values.
The error bars in Fig.~\ref{fig:fig2_resource_state_characterization}b correspond to a $99.7\%$ confidence level.

\subsection{Adaptive measurement basis choice} \label{method:compenstaion_measurement_byproduct}
\begin{figure}[t!]
\includegraphics{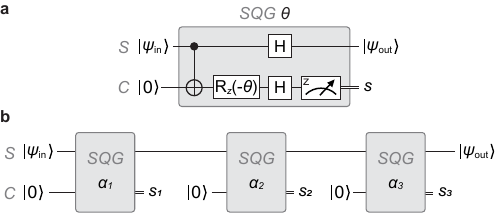}
\caption{\label{fig:fig6_measurement_basis_choice}{
\bf Principle of measurement-based single-qubit rotation.} 
 (a) An elementary circuit that realizes a measurement-based single-qubit rotation parameterized by the projection angle $\theta$. 
Passing an input qubit in state $|\psi_\mathrm{in}\rangle$ and an auxiliary qubit in state $|0\rangle$ through the sequence generates the output state $|\psi_\mathrm{out}\rangle$ accompanied by the binary measurement result $s$.
(b) Three measurement-based single-qubit rotation cells are concatenated to realize arbitrary single-qubit unitary parameterized by the Euler angles $(\alpha_1, \alpha_2, \alpha_3)$.
}
\end{figure}

We realize an elementary measurement-based single-qubit rotation circuit with entangling gates and mid-circuit measurements.
The circuits starts with a server qubit $S$ in an arbitrary initial state $|\psi_\mathrm{in}\rangle$ and the client qubit $C$ prepared in the ground state.
After being entangled with the server qubit with the CNOT gate, the client qubit is applied with the transformation unitary $U_\theta=HR_z(-\theta)$ and then read out in the computational basis. Given the binary measurement result $s\in\{0,1\}$, the output state $|\psi_\mathrm{out}\rangle$ is expressed as~\cite{Nielsen2006}, $|\psi_\mathrm{out}\rangle = X^sHR_z(\theta) |\psi_\mathrm{in}\rangle \, $, see Fig~\ref{fig:fig6_measurement_basis_choice}a

Concatenating three elementary circuits with measurements in bases $B(\alpha_1^*)$, $B(\alpha_2^*)$, and $B(\alpha_3^*)$, we realize a measurement-based arbitrary single-qubit unitary, see Fig.~\ref{fig:fig6_measurement_basis_choice}b. 
The output state is given by $|\psi_\mathrm{out}\rangle = X^{s_3}HR_z(\alpha^*_3) X^{s_2}HR_z(\alpha^*_2) X^{s_1}HR_z(\alpha^*_1) |\psi_\mathrm{in}\rangle \, $.
Here, $s_i$ is the result of each of the three measurement steps $(i=1,2,3)$. 
We rewrite the state as 
$|\psi_\mathrm{out}\rangle = X^{s_3}Z^{s_2}HR_z[(-1)^{s_2}\alpha_3^*] Z^{s_1} HR_z[(-1)^{s_1}\alpha_2^*]HR_z(\alpha_1^*) |\psi_\mathrm{in}\rangle$, 
using relations $R_z(\theta)X = XR_z(-\theta)$ and $HX = ZH$.
We adaptively choose the measurement bases $\alpha_1^* = \alpha_1$, $\alpha_2^* = (-1)^{s_1}\alpha_2$, $\alpha_3^* = (-1)^{s_2}\alpha_3 + \pi s_1$, and at the end of the sequence we apply the feedforward pulse $Z^{s_2}X^{s_3}$, deterministically generating the output state
$HR_z(\alpha_3) R_x(\alpha_2) R_z(\alpha_1) |\psi_\mathrm{in}\rangle$.

When performing a measurement-based quantum computation with $w\geq2$ cluster states, the commutation relations between the byproducts $X^s$ and the subsequent C$Z$ gates have to be considered~\cite{Raussendorf2001}.
In particular, the commutation was accounted for in the realization of the Deutsch-Jozsa algorithm discussed in the main text, with the results shown in Fig.~\ref{fig:fig4_three_qubit_protocol}.
In the two-qubit gate featured in Fig.~\ref{fig:fig3_measurement_based_gates}d, the C$Z$ gate entangling the server qubits is applied before the measurement-based rotations, so there are no commutators to be accounted for.

\subsection{Measuring the information leakage}\label{method:measuring_info_leakage}
The Holevo information $\chi$~\cite{Holevo1973} quantifies the maximum amount of classical information that can be extracted from an ensemble of quantum states.
We use this quantity to examine how much information could be extracted from the final state at the server.
For this, we execute the measurement-based single-qubit unitary sequence to prepare the client output state for $N=30$ distinct states on the Bloch sphere, as described in Appendix~\ref{app:arbitrary_single_qubit_gate}. 
Each unitary $j$ is parametrized by the three rotation angles $\bm{\alpha}_{j} = (\alpha_{j,1}, \alpha_{j,2}, \alpha_{j,3})$. All rotation angles form the set $\{\bm{\alpha}_j\}_{j=1,2,\dots,N}$. 
We characterize the server quantum state $\rho_j^\mathrm{s}$ at the end of each unitary with quantum state tomography. The Holevo information at the server is calculated as
\begin{equation}
    \chi(\rho^\mathrm{s}) = S(\rho^\mathrm{s}) - \frac{1}{N}\sum_{j=1}^N S(\rho^\mathrm{s}_j) \, ,
\end{equation}
where $S(\rho)= -\mathrm{Tr}(\rho \ln \rho)$ is the von-Neumann entropy and $\rho^\mathrm{s}=\frac{1}{N}\sum_{j=1}^N \rho_j^\mathrm{s}$ is the average density matrix.

According to Holevo's theorem, $\chi$ is upper-bounded by 1 bit for all single-qubit states~\cite{Holevo1973}. 
On the other hand, $\chi=0$ implies that no information is revealed about the qubit's state. 

\section*{Supplementary Information}
\appendix
\section{Sample and experimental setup}\label{app:sample_and_experimental_setup}

We perform the experiment on the modular superconducting quantum device described in Ref.~\cite{Dalton2025}. 
The device consists of two quantum modules, each hosting three transmon qubits. Every qubit is equipped with a dedicated drive line, flux line, and readout circuit. 
Within each module, the readout circuits are coupled to a common feedline, while the two modules are read out through separate lines. 
Inter-module quantum channels are realized with static qubit-qubit couplers across the modules~\cite{Norris2026}.

\begin{figure}[t!]
\includegraphics{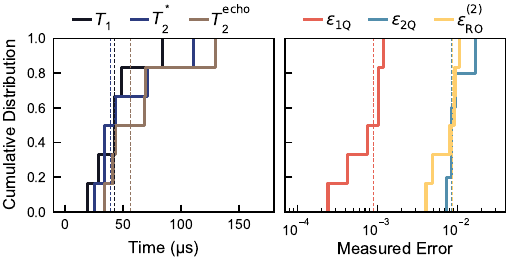}
\caption{\label{fig:fig7_device_characteristics}{\bf Device characteristics.} Cumulative distributions of (left panel) qubit relaxation time $T_1$ (black), dephasing time $T_2^*$ (blue), and Hahn-echo dephasing time $T_2^\mathrm{echo}$ (brown); (right panel) single-qubit gate (red), simultaneous two-qubit gate (cyan), and two-state readout (yellow) errors. Vertical dashed lines indicate the median values of the corresponding quantity.
}
\end{figure}

We characterize the coherence property of each qubit with standard methods~\cite{Schreier2008}. We extract a median relaxation time $T_1=43\,\mu\mathrm{s}$, dephasing time $T_2^*=39\,\mu\mathrm{s}$ and Hahn-echo dephasing time $T_2^\mathrm{echo}=56\,\mu\mathrm{s}$, see Fig.~\ref{fig:fig7_device_characteristics} (left panel).

In the experiment, we implement single-qubit gates with 48-ns microwave DRAG pulses~\cite{Motzoi2009} with Gaussian envelopes truncated at $\pm 2.5\sigma$, where $\sigma=\SI{9.6}{ns}$. The C$Z$ gates are realized with net-zero flux pulses~\cite{Rol2019, Negirneac2021} with an average duration of $\SI{60}{ns}$. We add $\SI{20}{ns}$ buffers before and after each flux pulse to account for the finite size of the pre-distortion filters~\cite{Hellings2025} and round each total C$Z$ gate time to an integer multiple of $\SI{8}{ns}$ to be commensurate with the granularity of the waveform playback of the control instruments. This results in an average $\SI{104}{ns}$ total duration of the C$Z$ gate. When reading out a qubit idled at its lower sweet spot, we apply a flux pulse to tune the qubit frequency closer to the readout resonator frequency, which dynamically increases the dispersive shift~\cite{Swiadek2024}.
We integrate the readout response for \SI{400}{ns} in the mid-circuit and final measurement rounds.

We benchmark the performance of the single-qubit gates with randomized benchmarking~\cite{Magesan2011, Epstein2014} and the performance of two-qubit gates with interleaved randomized benchmarking~\cite{Magesan2012, Corcoles2013, Barends2014}. We find a median single-qubit error $\varepsilon_\mathrm{1Q}=0.09\%$ and two-qubit error $\varepsilon_\mathrm{2Q}=0.8\%$, see Fig.~\ref{fig:fig7_device_characteristics} (right panel). We characterize the single-shot readout performance by preparing the qubits in either computational basis state $|0\rangle$ and $|1\rangle$ and determining the readout assignment probability matrix with single-shot readout. To initialize our qubit register, we perform a qubit readout before the start of the gate sequence and reject all instances where qubits are found in the excited state. The readout error $\varepsilon_\mathrm{RO}^{(2)}$ is calculated as the average of the misassignment probability $p_{0|1}$ and $p_{1|0}$, where $p_{a|b}$ stands for the probability of reading out the qubit in state $a$ when preparing in state $b$. We find a median readout error $\varepsilon_\mathrm{RO}^{(2)}=0.9\%$.

\begin{figure}[t!]
\includegraphics{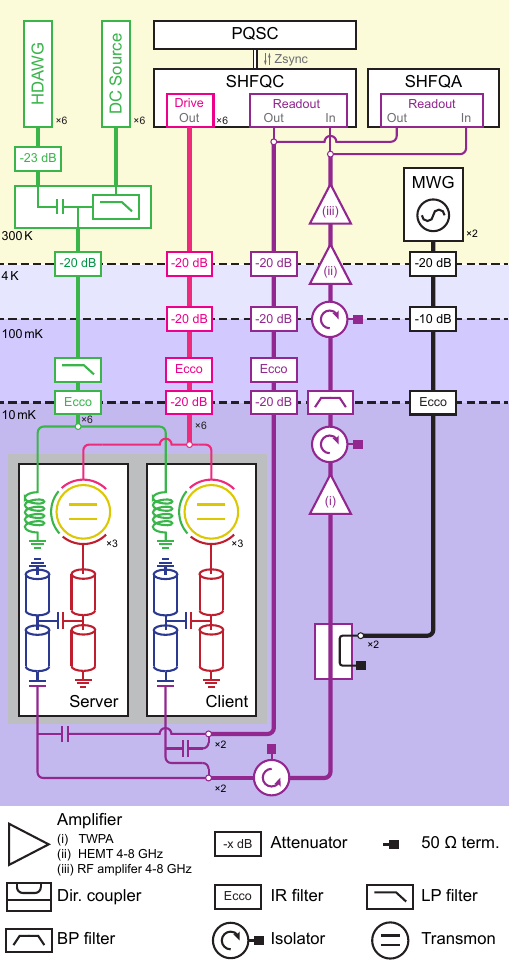}
\caption{\label{fig:fig8_setup_diagram} 
{\bf Schematic diagram of the experimental setup.} 
Flux lines (green), drive lines (pink), and readout lines (purple) transmit the signals between the control electronics and the device, including transmon qubits (yellow), readout resonators (red), and Purcell filters (blue). 
Thicker lines with number indicators illustrate multiple physical signal lines running in parallel. 
The background colors indicate the temperature stages of the experimental setup.
}
\end{figure}

The device is mounted to the \SI{10}{mK} stage of a dilution refrigerator in a standard wiring and shielding configuration~\cite{Krinner2019}. 
Room-temperature electronics are connected to the device through coaxial cables and microwave components, as illustrated in Fig.~\ref{fig:fig8_setup_diagram}.
Each qubit is biased by a DC source to set its idle operating frequency. 
Arbitrary waveform generators (HDAWGs) produce voltage pulses at a sampling rate of \SI{2.0}{GSa/s} to dynamically tune the qubit frequencies and implement two-qubit gates. 
The DC and pulsed voltage signals are combined with bias tees at room temperature. 
Drive channels of the qubit controller (SHFQC) generate microwave pulses at the qubit transition frequencies to implement single-qubit gates. 
The SHFQC readout output channel generates multiplexed signals at the readout resonator frequencies, which, after interacting with the client, are amplified by a chain consisting of a wide-band, near-quantum-limited traveling-wave parametric amplifier (TWPA)~\cite{Macklin2015}, a high-electron-mobility transistor (HEMT) amplifier, and a low-noise room-temperature amplifier.
The amplified signal is digitized and integrated by the SHFQC readout input channel. 
Following each measurement, the integrated results are compared with predefined thresholds to discriminate the qubit states. 
The thresholded outcomes are stored in the SHFQC and forwarded to the programmable quantum system controller (PQSC), which updates the subsequent reset and feedforward pulses.
To account for signal propagation delay and classical processing latency, a \SI{400}{ns} idle period is inserted between a mid-circuit readout pulse and the following client-qubit reset pulse. 
Finally, a quantum analyzer (SHFQA) is connected to the server quantum module. 
The SHFQA is not involved in executing the measurement-based quantum circuit but only used for preparing, calibrating, and verifying the blindness of the server. 

\section{Readout error mitigation}\label{app:readout_error_correction}
Readout of superconducting qubits is susceptible to energy relaxation and state-classification errors, which can cause the measured outcome to deviate from the ideally projected quantum state. 
Such readout errors may bias experimentally inferred quantities, and the introduced error usually scales with the number of readout involved in the experiment.
Despite that an individual two-state readout error in the experimental device is usually below $1\%$, readout error mitigation becomes crucial, for instance, when evaluating the cluster state fidelity, as up to 22 measurements are used to evaluate a stabilizer.
In less crucial cases, experimentally measured observables will also benefit from readout error mitigation to avoid the associated bias.
Here, we mitigate readout errors using information obtained from dedicated readout benchmarking.

We first characterize the readout assignment matrix $P_k$ for each qubit $k$, following the procedure described in Appendix~\ref{app:sample_and_experimental_setup}, which is given by
\begin{equation}
    P_k = \begin{pmatrix}
    p_{0|0} & p_{1|0} \\
    p_{0|1} & p_{1|1}
    \end{pmatrix}  \, .
\end{equation}
On the experimental device, each readout resonator is equipped with a dedicated Purcell filter. 
This design suppresses readout-induced crosstalk, such that the correlated readout error is negligible compared to other error sources~\cite{Heinsoo2018}. 
Under this assumption, the readout assignment matrix for a $n$-qubit register is well approximated by the tensor product $P = P_1 \otimes P_2 \otimes \cdots P_n$. 

In this work, all directly-measured observables $O$ that require readout-error mitigation can be written as tensor products of single-qubit operators, $O = O_1 \otimes O_2 \otimes \cdots \otimes O_n$.
We repeat the measurement for $M$ shots to estimate the expectation value $\langle O \rangle$, obtaining a set of measured bitstrings $|s^j\rangle = |s^j_1 s^j_2 \cdots s^j_n\rangle$, with $j = 1, 2, \dots, M$.
Assuming uncorrelated readout errors, an unbiased estimator for $\langle O \rangle$ is given by~\cite{Bravyi2021}
\begin{equation}
    \langle O \rangle = \frac{1}{M} \sum_{j=1}^{M} \prod_{k=1}^{n} \langle e | O_k P_k^{-1} |s_k^j\rangle \, .
\end{equation}
where $|e\rangle = |0\rangle + |1\rangle$ is an auxiliary vector introduced for simplifying the notation.

\section{Numerical simulation of the measurement-based quantum computation sequences} \label{app:numerical_simulation_of_mbqc_circuit}

We simulate the performance of the demonstrated measurement-based quantum computation sequences with circuit-level simulation using Qiskit~\cite{Javadi-Abhari2024}. 
We setup the simulation according to the measurement patterns and the required physical operations, see Appendix~\ref{app:gate_sequence_and_time_budget}, and take into account the qubit idling time while quantum gates are applied to other qubits or while the feedforward control is being processed by the classical electronics. 
When a qubit is involved in a single-, two-qubit gate or a readout, we apply a symmetric depolarizing channel~\cite{Nielsen2010}:
\begin{equation}
    \rho \mapsto (1-p)\rho + p\frac{I}{2^N} \, ,
\end{equation}
where $N$ is the number of qubits involved in the operation and the depolarizing parameter $p$ is chosen to yield the same error rate as extracted from separate characterization measurements, see Appendix.~\ref{app:sample_and_experimental_setup}. 
Besides, when a qubit is idled for time $\tau$, we introduce a thermal relaxation channel, defined as 
\begin{equation}
    \begin{pmatrix}
    \rho_{00} & \rho_{01} \\
    \rho_{10} & \rho_{11}
    \end{pmatrix} \mapsto
    \begin{pmatrix}
    \rho_{00} + \rho_{11}(1-e^{-\tau/T_1}) & \rho_{01}e^{-\tau/T_2} \\
    \rho_{10}e^{-\tau/T_2} & \rho_{11}e^{-\tau/T_1}
    \end{pmatrix} \, ,
\end{equation}
where $T_1, T_2$ are the energy relaxation time and dephasing time of the qubit. With this model, we simulate the output state fidelities featured in Fig.~\ref{fig:fig2_resource_state_characterization},  Fig.~\ref{fig:fig3_measurement_based_gates} and Fig.~\ref{fig:fig4_three_qubit_protocol}. 
The quantitative agreement between the measurement and simulation shows that we have correctly captured the major noise sources in the sequence.

Using the simulation framework, we analyze the error composition of the demonstrated measurement-based quantum computation sequences with the method presented in Ref.~\cite{Chen2021p}. 
First, we simulate the measurement-based quantum computation sequence with error rates and coherence properties extracted from measurements performed on the experimental device. 
Then, we halve all errors in each source $i$, $i \in \{\text{single-qubit gate, two-qubit gate, readout, idling}\}$ in turn to estimate the partial derivative of the total error rate with respect to the mean error rate of the source:
\begin{equation}
    \frac{\partial E(\bm{p}_i)}{\partial \bar p_i} \approx \frac{E(\bm{p}_i) - E(\bm{p}_i / 2)}{\bar p_i - (\bar p_i / 2)} \, ,
\end{equation}
where $\bm{p}_i = (p_{i,1}, p_{i,2}, \dots)$ is the set of all errors from source $i$, $\bar p_i$ is the mean of $\bm{p}_i$, and $E=1-F$ is the error of the output state. We take the weighted fraction
\begin{equation}
    X_i = \frac{\bar p_i \frac{\partial E(\bm{p}_i)}{\partial \bar p_i} }{\sum_i \bar p_i \frac{\partial E(\bm{p}_i)}{\partial \bar p_i}} \, ,
\end{equation}
to be the participation of the source $i$ in the output state error.

We simulate the error composition of the measurement-based T-gate and the maximally entangled state, as well as the measurement-based Deutsch–Jozsa algorithm presented in the main text. 
The corresponding results are summarized in Table~\ref{tab:error_composition}. 
Across all cases, we find that idling accounts for roughly half of the total error, making it the dominant error source. 
The measurement-based two-qubit gate sequence exhibits the highest fraction of idling error, primarily due to the serialized execution of two-qubit gates. 
In contrast, for the measurement-based Deutsch–Jozsa algorithm, the parallelization of two-qubit gates $(\mathrm{S}_1, \mathrm{C}_1)$ and $(\mathrm{S}_3, \mathrm{C}_3)$ reduces the relative contribution from idling. 
Two-qubit gates constitute the next major error source, contributing about 30\% of the total error, due to their relatively large count and higher intrinsic gate error. 
Single-qubit gates and readout together contribute approximately 10\%, with a decreasing relative contribution as the computational sequence becomes more complex.

\begin{table}[t!]
\centering
\begin{tabular}{lccc}
\toprule
\textbf{Error Source} & \multicolumn{3}{c}{\bf Measurement-based sequence} \\
 & Single-qubit &  Two-qubit & Deutsch-Jozsa \\
 
\midrule
Single-qubit gate & $11\%$ & $ 7\%$ & $ 7\%$ \\
C$Z$ gate           & $26\%$ & $30\%$ & $32\%$ \\
Readout           & $16\%$ & $ 9\%$ & $14\%$ \\
Idling            & $47\%$ & $54\%$ & $47\%$ \\
\bottomrule
\end{tabular}
\caption{\label{tab:error_composition} Simulated error decomposition of the measurement-based single- and two-qubit gates and the measurement-based Deutsch-Jozsa algorithm.}
\vspace{-1em}
\end{table}

\section{Gate sequence and time budget}\label{app:gate_sequence_and_time_budget}

\begin{figure}[b]\setlength{\hfuzz}{1.1\columnwidth}
\begin{minipage}{\textwidth}
\includegraphics[width=\textwidth]{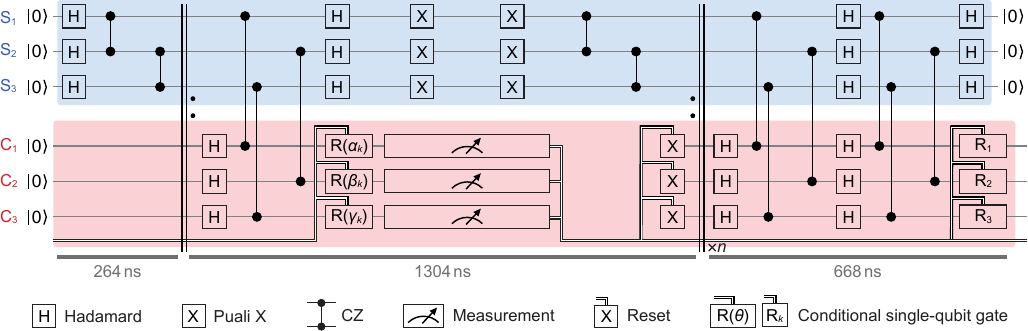}
\caption{\label{fig:fig9_timing_diagram}{\bf Timing diagram.} Decomposing $w=3$ measurement-based quantum computation circuit with native operations of the modular device. The time budget of each step is shown below the circuit.
$S_j$ and $C_j,\, (j=1,2,3)$ denotes the qubits on the server and client module, respectively.
}
\end{minipage}
\end{figure}

We realize the measurement-based quantum computation sequence with the native gate set in our architecture. 
We implement single-qubit $x$ and $y$ rotations with 48-ns DRAG pulses~\cite{Motzoi2009}. We realize single-qubit $z$ rotations with virtual $Z$ gates~\cite{McKay2017}. 
We decompose CNOT gates into 104-ns C$Z$ gates~\cite{Rol2019, Negirneac2021} and $y$ rotations. 
Qubits are read out with simultaneous, frequency-multiplexed readout tones up to $\SI{400}{ns}$.
Before each measurement run, we read out every qubit and herald all qubits in the $|0\rangle$ state~\cite{Riste2012, Johnson2012}.
When applicable, multiple single-qubit gates are replaced with a single equivalent gate to reduce the sequence length. 
During the readout on the client qubits, we apply two Carr-Purcell-Meiboom-Gill dynamical decoupling pulses~\cite{Carr1954, Meiboom1958} on every server qubit to mitigate low-frequency noise,
see Fig.~\ref{fig:fig9_timing_diagram} for a decomposed gate set of width $w=3$ measurement-based quantum computation sequences. 
For sequences with $w=1$ or $2$, we leave the unused qubits in the ground state.
We parallelize the execution of single-qubit gates, as well as that of two-qubit gates $\mathrm{S}_1$-$\mathrm{C}_1$, $\mathrm{S}_3$-$\mathrm{C}_3$. 
Here, $\mathrm{S}_{j}$ and $\mathrm{C}_{j}$ denote the $j$-th qubit on the server and the client module, respectively.
As the static couplers on the device generates an always-on coupling, all other two-qubit gates have to be executed sequentially to avoid interactions with the qubits not involved in the gate.

With the described decomposition applied to $w=3$ measurement-based quantum computation sequences, the initial preparation of the cluster state on the server module takes $\SI{264}{ns}$. 
Each cycle of cluster state preparation, transfer, and measurement takes $\SI{1304}{ns}$. 
The final SWAP circuit takes $\SI{668}{ns}$. 
A 400-ns idling time is reserved before the reset pulse at the end of each cycle.
The idling time consists of a 300-ns signal processing delay in the classical electronics, a 85-ns round-trip signal propagation delay between the quantum device and the electronics, and a 15-ns buffer time.
When the classical information is processed, \emph{e.g.}, for the client qubit reset or for generating the adaptive basis rotation pulse, quantum sequences unconditioned on the processing result are executed in parallel.
The co-processing of classical and quantum information further reduced the execution time and enables an efficient implementation of the experimental sequence.
\clearpage

\section{Measurement-based arbitrary single-qubit unitary} \label{app:arbitrary_single_qubit_gate}

Beyond the T-gate, we also demonstrate the ability to perform arbitrary single-qubit unitaries within the measurement-based framework.

To this aim, we sweep one of the three Euler angles, $\alpha_1, \alpha_2$ and $\alpha_3$, while keeping the other two fixed.
Specifically, we choose to vary either $\alpha_1$ or $\alpha_2$ in two separate experiments, while keeping the other constant and setting $\alpha_3=0$.
First, we fix $\alpha_2=\alpha_3=0$ and sweep $\alpha_1$ within $[0, 2\pi)$, essentially varying the excitation probability of the output state. 
We characterize each output state using quantum state tomography and estimate the density matrix $\rho$ of the experimental output state. 
We compare $\rho$ with the ideal output state $| \psi_\mathrm{ideal} \rangle$ and calculate the output state fidelity $F = \langle \psi_\mathrm{ideal} | \rho | \psi_\mathrm{ideal} \rangle$ as well as the expectation value of the Pauli operators $\langle O \rangle=\mathrm{Tr}(O\rho), O\in\{X,Y,Z\}$. 

\begin{figure}[t!]
\includegraphics{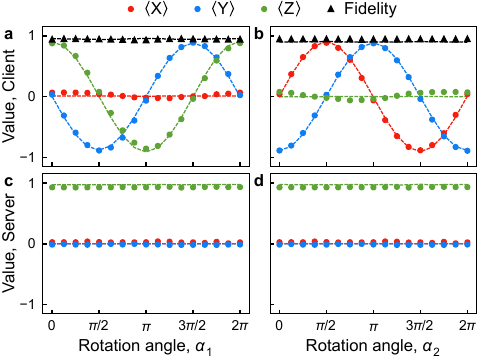}
\caption{\label{fig:fig10_parametric_angle_sweep}{
\bf Measurement-based preparation of single-qubit states over the Bloch sphere.} 
Measured output state fidelity (black triangles) and expectation value (circles) of Pauli X (blue), Y (red), and Z (green) operators of single-qubit states at the end of the sequence. 
Dashed lines shows the result from circuit-based simulations.
The polar angle sweep is implemented by sweeping $\alpha_1$ and fixing $\alpha_2=\alpha_3=0$; the corresponding server and client states are shown in {\bf (a)} and {\bf (c)}, respectively.
The azimuthal angle sweep is implemented by sweeping $\alpha_2$ and fixing $\alpha_1=\pi/2, \alpha_3=0$; the corresponding server and client states are shown in {\bf (b)} and {\bf (d)}, respectively.
}
\end{figure}

For the sweep over $\alpha_1$, we extract an average output state fidelity $F = 0.939(6)$. 
We find that $\langle Y\rangle$ and $\langle Z\rangle$ follow sinusoidal oscillations as expected, with a contrast of 0.892(6) and 0.867(6), respectively, see Fig.~\ref{fig:fig10_parametric_angle_sweep}a.
We calculate the purity, $\gamma \equiv Tr(\rho^2)$, and extract an average coherent error $\varepsilon_\mathrm{coh} = (1+\gamma) / 2 - F = 0.005(1)$ for the sweep, suggesting that the gate sequence is dominated by incoherent errors. 
Similarly, we fix $\alpha_1=-\pi / 2, \alpha_3 = 0$ and vary $\alpha_2$, preparing a uniform superposition between the computational basis states, $|0\rangle$ and $|1\rangle$, with a varying relative phase between them.
For this sweep, we measure an average output state fidelity $F=0.941(2)$, as well as a contrast of 0.883(3) and 0.881(2), respectively, from the sinusoidal oscillation of $\langle X \rangle$ and $\langle Y \rangle$, see Fig.~\ref{fig:fig10_parametric_angle_sweep}b.
The coherent error is found to be 0.0046(6), in line with the low coherent error in the parametric sweep over $\alpha_1$.
The final state of the server qubits shows no dependence to the rotation angles, see Fig.~\ref{fig:fig10_parametric_angle_sweep}c,~d, showing blindness to the executed computing sequence. 
We do not observe any systematic dependence of the server-qubit state on the rotation angles, and therefore attribute the residual fluctuations to statistical uncertainty in the quantum state tomography due to the finite number of measurement shots, as well as to time-dependent noise sources in the device, such as fluctuations of two-level systems.

\begin{figure*}[t!]
    \includegraphics{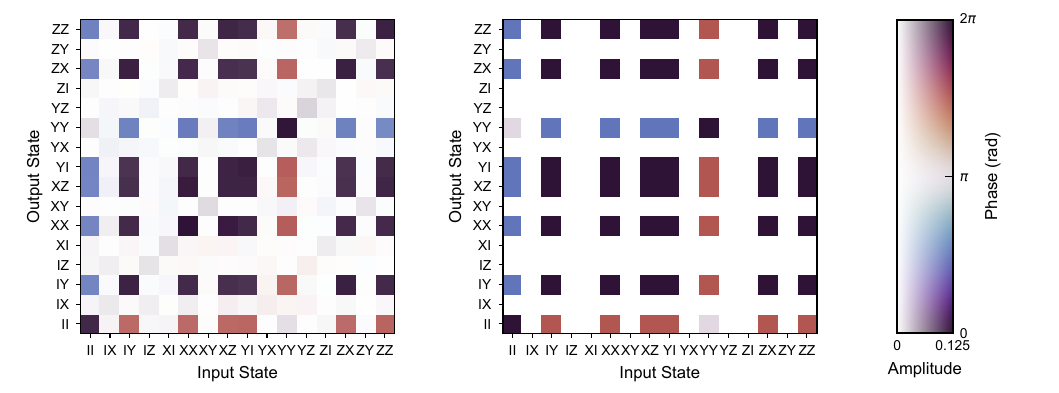}
    \caption{\label{fig:fig11_measurement_based_entangled_state_process_tomography} {\bf Process tomography of the measurement-based maximally entangled state generation sequence.} Measured (left) and ideal (right) process matrix $\chi_\mathrm{2q}$. Colors displays the amplitude and phase of each matrix element, as shown in the color bar.} 
\end{figure*}

\section{Quantum process tomography of the measurement-based single-and two-qubit gates}\label{app:quantum_process_tomography}

We use the process matrix formalism to characterize the quantum processes implemented by the measurement-based quantum computation sequences.
Within this formalism, a quantum process denoted by a superoperator $\sigma$, acting on an $n$-qubit density matrix $\rho$, can be written as
\begin{equation}
    \sigma(\rho) = \sum_{j,k} \varepsilon_{jk} A_j \rho A_k^\dagger \, ,
\end{equation}
where $A_j, A_k \in \{I, X, Y, Z\}^{\otimes n}$ form a complete basis of $n$-qubit Pauli operators, and $\varepsilon$ is the process matrix.

Experimentally, we reconstruct $\varepsilon$ by preparing a set of input states $\{\rho_\mathrm{in}\}$ and characterizing their corresponding output states $\{\rho_\mathrm{out}\}$. The input states are generated by applying initial rotation pulses
$R_\mathrm{start} \in \{I, R_x(\pi), R_y(\pi/2), R_x(\pi/2)\}^{\otimes n}$
to the server qubits prior to the measurement-based quantum computation sequence, preparing
\begin{equation}
    \rho_\mathrm{in}
    =
    R_\mathrm{start}
    (|0\rangle\langle 0|)^{\otimes n}
    R_\mathrm{start}^\dagger \, .
\end{equation}
For each prepared input state, we apply final rotation pulses
$R_\mathrm{end} \in \{I, R_x(\pi), R_y(\pi/2), R_x(\pi/2)\}^{\otimes n}$
and perform measurements on the client qubits to estimate the output state $\rho_\mathrm{out}$ using maximum-likelihood estimation~\cite{James2001}. The process matrix $\varepsilon$ is then reconstructed by linear inversion from the experimentally prepared input states ${\rho_\mathrm{in}}$ and the corresponding measured output states ${\rho_\mathrm{out}}$.

We characterize the measurement-based T-gate process, with the reconstructed process matrix shown in Fig.~\ref{fig:fig3_measurement_based_gates}a of the main text. 
We also perform process tomography on the measurement-based maximally entangled state generation sequence discussed in Fig.~\ref{fig:fig3_measurement_based_gates}c of the main text, with the corresponding results presented in Fig.~\ref{fig:fig11_measurement_based_entangled_state_process_tomography}. 
We extract a 0.902 and 0.845 process fidelity from the two sequences, respectively.
In both cases, we observe that the residual errors mostly appear on the diagonal elements of the process matrices.
This behavior arises because the random mid-circuit projective measurements effectively implement Pauli twirling, which depolarizes coherent errors. 
Errors on the off-diagonal elements are still visible and are most likely attributable to coherent errors in the single- and two-qubit gates within the SWAP circuit, as well as the residual $ZZ$ interactions between the involved physical qubits.

\section{Performance and resource analysis of the measurement-based quantum computation sequences}\label{app:performance_and_overhead_analysis_mbqc}
\subsection{Measurement-based single- and two-qubit gates}

\begin{table}[b!]
\centering
\begin{tabular}{lcc}
\toprule

 & \multicolumn{2}{c}{\bf Measurement-based sequences} \\
 Count & 1Q Unitary &  2Q Entanglement \\
\midrule
Stationary qubit & $2$ & $4$ \\
Single-qubit gate & $24$ & $24$ \\
Two-qubit gate  & $5$ & $7$ \\
Mid-circuit readout  & $3$ & $4$ \\
Feedforward idling  & $3$ & $2$ \\
\bottomrule
\end{tabular}
\caption{\label{tab:error_source_count_mbqc_gates} Resource count for implementing a measurement-based (MB) arbitrary single-qubit unitary and a maximally entangled two-qubit state.}
\end{table}
First, we compare the output state fidelity of blind quantum computation sequences with the expected performance of gate-based sequences on the experimental device. 
In Table~\ref{tab:error_source_count_mbqc_gates}, we present the physical resources required for implementing a measurement-based arbitrary single-qubit unitary and a maximally entangled two-qubit state.
The overhead is higher than the implementation with the gate-based circuits, thus improvements needs to be made in measurement-based quantum computation to be comparable with the performance of the gate-based circuit.
From Fig~\ref{fig:fig3_measurement_based_gates}b, we extract an average output state fidelity of $94\%$ from the parametric sweep over the polar and azimuthal angles. 
In comparison, a microwave-activated arbitrary single-qubit unitary requires three successive microwave gates, where each gate has an average infidelity of $0.086\%$ on the experimental device.
We thus infer that with an improvement factor of 24 (i.e. all individual errors get reduced to their 1/24), the measurement-based quantum computation circuit will yield the same fidelity as the gate-based single-qubit unitary.
Similarly, an improvement factor of 10 is required for the measurement-based quantum computation circuit to yield comparable output state fidelity as the gate-based generation of maximally entangled state.
In Section~\ref{app:performance_and_overhead_analysis_mbqc}3, we show that a comparable improvement factor is within reach of state-of-the-art superconducting technology, enabling blind quantum computation with a practical overhead.

\subsection{Measurement-based Deutsch-Jozsa algorithm}
Here, we compare the overhead of the measurement-based and gate-based circuits for realizing the Deutsch-Jozsa algorithm. 
To analyze the resource requirements scaling with input size, $n$, we focus on a balanced oracle, whose circuit implementation consists of a series of CNOT gates between the data and the auxiliary qubit~\cite{Tame2010}. 
We count the resources required for the gate-based and measurement-based circuits, see Table~\ref{tab:error_source_count}.

\begin{table}[t!]
\centering
\begin{tabular}{lcc}
\toprule
 Count &  Gate-based &  MBQC \\
\midrule
Physical qubit & $n+1$ & $2n+2$ \\
Single-qubit gate & $4n+1$ & $18n$ \\
Two-qubit gate  & $n$ & $4n+5$ \\
Mid-circuit readout  & $0$ & $2n+2$ \\
Feedforward idling  & $0$ & $4n+4$ \\
\bottomrule
\end{tabular}
\caption{\label{tab:error_source_count} Resource count for implementing Deutch-Jozsa algorithm with gate-based and measurement-based quantum computation (MBQC) circuit. $n$ represents the number of the input qubit of an oracle.}
\end{table}

From the experimental device, we extract an average single-qubit gate error $\bar\epsilon_\mathrm{1q} = 0.086\%$, an average two-qubit gate error $\bar\epsilon_\mathrm{2q} = 1.02\%$, an average feedback time idling error $\bar\epsilon_\mathrm{idle} = 0.96\%$, and an average mid-cricuit readout error $\bar\epsilon_\mathrm{ro} = 0.77\%$. 
We use a pass-fail model, estimating the success rate:
\begin{equation}
F = (1-\bar\epsilon_\mathrm{1q})^{n_\mathrm{1q}} * (1-\bar\epsilon_\mathrm{2q})^{n_\mathrm{2q}} * (1-\bar\epsilon_\mathrm{idle})^{n_\mathrm{idle}} * (1-\bar\epsilon_\mathrm{ro})^{n_\mathrm{ro}}
\end{equation}
of the algorithm, where $n_\mathrm{1q},n_\mathrm{2q},n_\mathrm{idle},n_\mathrm{ro}$ are the occurrence of the error sources, correspondingly. 
Using the resource count from Table~\ref{tab:error_source_count}, we obtain the success probability of the Deutsch-Jozsa algorithm for an n-qubit query..
We estimate that with an improvement of all error sources by a factor of 7.7, the measurement-based circuit will have a comparable performance as the gate-based circuit at an input size of $n=20$, and an asymptotically lower improvement factor is needed at larger $n$.

\subsection{Performance estimate of blind quantum computation using superconducting devices with state-of-the-art fidelities}

As the blind quantum computation protocol introduced here can be deployed on a powerful quantum server, we project its performance onto state-of-the-art superconducting quantum devices, thereby estimating the performance achievable with current and near-term technology.
We assume an error rate of $1\times10^{-4}$ per single-qubit gate~\cite{Somoroff2023b, Li2023n, Rower2024, Marxer2025}, $1\times10^{-3}$ per two-qubit gate~\cite{Negirneac2021, Ding2023, Lin2024a, Li2024c, Marxer2025, IBM2025}, and $1\times10^{-3}$ per readout~\cite{Swiadek2024, Spring2025, Wang2025r}, together with relaxation and dephasing times of $\SI{1}{ms}$ for $T_1$ and $T_2$~\cite{Place2021, Wang2022a, Somoroff2023b, Tuokkola2025, Wang2025g, Bland2025}.
Based on these parameters, we simulate the output-state fidelities of the measurement-based T gate and the maximally entangled state, as well as the measurement-based Deutsch–Jozsa algorithm described in the main text.
With these state-of-the-art performance levels, the measurement-based T gate and maximally entangled state are expected to achieve output-state fidelities above 0.99, while the measurement-based Deutsch–Jozsa algorithm yields a correct output with a probability exceeding 0.98.
These results highlight the feasibility of the proposed protocol with currently accessible technology. 
Note that the overhead of blind quantum computation is calculated assuming that all measurement-based single-qubit rotations are kept blind to the server.
If only a subset of the rotations needs to be blind, the overhead can be reduced by applying the rest of the rotations directly on the server~\cite{Poshtvan2025, Baranes2025a}.

\begin{table}[t!]
\centering
\begin{tabular}{lccc}
\toprule
 & \multicolumn{3}{c}{\bf Blind sequences} \\
 & 1Q Unitary & 2Q Entanglement & Deutsch-Jozsa \\
\midrule
This experiment   & 0.942 & 0.849 & 0.797 \\
Projection  & 0.992 & 0.995 & 0.984 \\
\bottomrule
\end{tabular}
\caption{\label{tab:fidelity_improvement_prediction} Output state fidelity of the measurement-based sequences from this experiment compared to the projection using the state-of-the-art error probabilities, details are discussed in the text.}
\end{table}

Further improvements could be achieved by incorporating error correction into the blind quantum computing framework—for example, by extending the client–server interface to a two-dimensional grid and employing a three-dimensional cluster state as a fault-tolerant resource for blind quantum computation~\cite{Raussendorf2006, Raussendorf2007a}.

\bibliographystyle{apsrev4-2-title-etal}

\begin{thebibliography}{92}%
\makeatletter
\providecommand \@ifxundefined [1]{%
 \@ifx{#1\undefined}
}%
\providecommand \@ifnum [1]{%
 \ifnum #1\expandafter \@firstoftwo
 \else \expandafter \@secondoftwo
 \fi
}%
\providecommand \@ifx [1]{%
 \ifx #1\expandafter \@firstoftwo
 \else \expandafter \@secondoftwo
 \fi
}%
\providecommand \natexlab [1]{#1}%
\providecommand \enquote  [1]{``#1''}%
\providecommand \bibnamefont  [1]{#1}%
\providecommand \bibfnamefont [1]{#1}%
\providecommand \citenamefont [1]{#1}%
\providecommand \href@noop [0]{\@secondoftwo}%
\providecommand \href [0]{\begingroup \@sanitize@url \@href}%
\providecommand \@href[1]{\@@startlink{#1}\@@href}%
\providecommand \@@href[1]{\endgroup#1\@@endlink}%
\providecommand \@sanitize@url [0]{\catcode `\\12\catcode `\$12\catcode `\&12\catcode `\#12\catcode `\^12\catcode `\_12\catcode `\%12\relax}%
\providecommand \@@startlink[1]{}%
\providecommand \@@endlink[0]{}%
\providecommand \url  [0]{\begingroup\@sanitize@url \@url }%
\providecommand \@url [1]{\endgroup\@href {#1}{\urlprefix }}%
\providecommand \urlprefix  [0]{URL }%
\providecommand \Eprint [0]{\href }%
\providecommand \doibase [0]{https://doi.org/}%
\providecommand \selectlanguage [0]{\@gobble}%
\providecommand \bibinfo  [0]{\@secondoftwo}%
\providecommand \bibfield  [0]{\@secondoftwo}%
\providecommand \translation [1]{[#1]}%
\providecommand \BibitemOpen [0]{}%
\providecommand \bibitemStop [0]{}%
\providecommand \bibitemNoStop [0]{.\EOS\space}%
\providecommand \EOS [0]{\spacefactor3000\relax}%
\providecommand \BibitemShut  [1]{\csname bibitem#1\endcsname}%
\let\auto@bib@innerbib\@empty
\bibitem [{\citenamefont {Gidney}(2025)Gidney, C.}]{Gidney2025b}%
  \BibitemOpen
  \bibfield  {author} {\bibinfo {author} {Gidney, C.},\ }\bibfield  {title} {\bibinfo {title} {How to factor 2048 bit {RSA} integers with less than a million noisy qubits},\ }\href {https://arxiv.org/abs/2505.15917} {\bibfield  {journal} {\bibinfo  {journal} {arXiv:2505.15917}\ } (\bibinfo {year} {2025})}\BibitemShut {NoStop}%
\bibitem [{\citenamefont {Low}\ and\ \citenamefont {Chuang}(2019)Low, Guang Hao and Chuang, Isaac L}]{Low2019a}%
  \BibitemOpen
  \bibfield  {author} {\bibinfo {author} {Low, G.~H.}\ and\ \bibinfo {author} {Chuang, I.~L.},\ }\bibfield  {title} {\bibinfo {title} {Hamiltonian simulation by qubitization},\ }\href {https://doi.org/https://doi.org/10.22331/q-2019-07-12-163} {\bibfield  {journal} {\bibinfo  {journal} {Quantum}\ }\textbf {\bibinfo {volume} {3}},\ \bibinfo {pages} {163} (\bibinfo {year} {2019})}\BibitemShut {NoStop}%
\bibitem [{\citenamefont {Berry}\ \emph {et~al.}(2024)Berry, Dominic W. and Su, Yuan and Gyurik, Casper and King, Robbie and Basso, Joao and Barba, Alexander Del Toro and Rajput, Abhishek and Wiebe, Nathan and Dunjko, Vedran and Babbush, Ryan}]{Berry2024a}%
  \BibitemOpen
  \bibfield  {author} {\bibinfo {author} {Berry, D.~W.}, \bibinfo {author} {Su, Y.}, \bibinfo {author} {Gyurik, C.}, \emph {et~al.},\ }\bibfield  {title} {\bibinfo {title} {Analyzing prospects for quantum advantage in topological data analysis},\ }\href {https://doi.org/10.1103/PRXQuantum.5.010319} {\bibfield  {journal} {\bibinfo  {journal} {PRX Quantum}\ }\textbf {\bibinfo {volume} {5}},\ \bibinfo {pages} {010319} (\bibinfo {year} {2024})}\BibitemShut {NoStop}%
\bibitem [{\citenamefont {Ransford}\ \emph {et~al.}(2025)Ransford, A. and Allman, M. S. and Arkinstall, J. and Campora III, J. P. and Cooper, S. F. and Delaney, R. D. and Dreiling, J. M. and Estey, B. and Figgatt, C. and Hall, A. and Husain, A. A. and Isanaka, A. and Kennedy, C. J. and Kotibhaskar, N. and Madjarov, I. S. and Mayer, K. and Milne, A. R. and Park, A. J. and Reed, A. P. and Ancona, R. and Andersen, M. P. and Andres-Martinez, P. and Angenent, W. and Argueta, L. and Arkin, B. and Ascarrunz, L. and Baker, W. and Barnes, C. and Bartolotta, J. and Berg, J. and Besand, R. and Bjork, B. and Blain, M. and Blanchard, P. and Blume-Kohout, R. and Bohn, M. and Borgna, A. and Botamanenko, D. Y. and Boutelle, R. and Brown, N. and Buckingham, G. T. and Burdick, N. Q. and Burton, W. C. and Carey, V. and Carron, C. J. and Chambers, J. and Children, J. and Colussi, V. E. and Crepinsek, S. and Cureton, A. and Davies, J. and Davis, D. and DeCross, M. and Deen, D. and Delaney, C. and DelVento, D. and DeSalvo, B. J.
  and Dominy, J. and Duncan, R. and Eccles, V. and Edgington, A. and Erickson, N. and Erickson, S. and Ertsgaard, C. T. and Evans, B. and Evans, T. and Fabrikant, M. I. and Fischer, A. and Foltz, C. and Foss-Feig, M. and Francois, D. and Freyberg, B. and Gao, C. and Garay, R. and Garvin, J. and Gaudiosi, D. M. and Gilbreth, C. N. and Giles, J. and Glynn, E. and Graves, J. and Hansen, A. and Hayes, D. and Heidemann, L. and Higashi, B. and Hilbun, T. and Hines, J. and Hlavaty, A. and Hoffman, K. and Hoffman, I. M. and Holliman, C. and Hooper, I. and Horning, B. and Hostetter, J. and Hothem, D. and Houlton, J. and Hout, J. and Hutson, R. and Jacobs, R. T. and Jacobs, T. and Johannsen, M. and Johansen, J. and Jones, L. and Julian, S. and Jung, R. and Keay, A. and Klein, T. and Koch, M. and Kondo, R. and Kong, C. and Kosto, A. and Lawrence, A. and Liefer, D. and Lollie, M. and Lucchetti, D. and Lysne, N. K. and Lytle, C. and MacPherson, C. and Malm, A. and Mather, S. and Mathewson, B. and Maxwell, D. and
  McCaffrey, L. and McDougall, H. and Mendoza, R. and Mills, M. and Morrison, R. and Narmour, L. and Nguyen, N. and Nugent, L. and Olson, S. and Ouellette, D. and Parks, J. and Peters, Z. and Petricka, J. and Pino, J. M. and Polito, F. and Preidl, M. and Price, G. and Proctor, T. and Pugh, M. and Ratcliff, N. and Raymondson, D. and Rhodes, P. and Roman, C. and Roy, C. and Ryan-Anderson, C. and Sanchez, F. B. and Sangiolo, G. and Sawadski, T. and Schaffer, A. and Schow, P. and Sedlacek, J. and Semenenko, H. and Shevchuk, P. and Shore, S. and Siegfried, P. and Singhal, K. and Sivarajah, S. and Skripka, T. and Sletten, L. and Spaun, B. and Sprenkle, R. T. and Stoufer, P. and Tader, M. and Taylor, S. F. and Thompson, T. H. and Tobey, R. and Tran, A. and Tran, T. and Vittorini, G. and Volin, C. and Walker, J. and White, S. and Wilson, D. and Wolf, Q. and Wringe, C. and Young, K. and Zheng, J. and Zuraski, K. and Baldwin, C. H. and Chernoguzov, A. and Gaebler, J. P. and Sanders, S. J. and Neyenhuis, B. and Stutz,
  R. and Bohnet, J. G.}]{Ransford2025}%
  \BibitemOpen
  \bibfield  {author} {\bibinfo {author} {Ransford, A.}, \bibinfo {author} {Allman, M.~S.}, \bibinfo {author} {Arkinstall, J.}, \emph {et~al.},\ }\bibfield  {title} {\bibinfo {title} {Helios: A 98-qubit trapped-ion quantum computer},\ }\href {https://arxiv.org/abs/2511.05465} {\bibfield  {journal} {\bibinfo  {journal} {arXiv:2511.05465}\ } (\bibinfo {year} {2025})}\BibitemShut {NoStop}%
\bibitem [{\citenamefont {Chiu}\ \emph {et~al.}(2025)Neng-Chun Chiu and Elias C. Trapp and Jinen Guo and Mohamed H. Abobeih and Luke M. Stewart and Simon Hollerith and Pavel Stroganov and Marcin Kalinowski and Alexandra A. Geim and Simon J. Evered and Sophie H. Li and Lisa M. Peters and Dolev Bluvstein and Tout T. Wang and Markus Greiner and Vladan Vuletić and Mikhail D. Lukin}]{Chiu2025a}%
  \BibitemOpen
  \bibfield  {author} {\bibinfo {author} {Chiu, N.-C.}, \bibinfo {author} {Trapp, E.~C.}, \bibinfo {author} {Guo, J.}, \emph {et~al.},\ }\bibfield  {title} {\bibinfo {title} {Continuous operation of a coherent 3,000-qubit system},\ }\href {https://doi.org/https://doi.org/10.1038/s41586-025-09596-6} {\bibfield  {journal} {\bibinfo  {journal} {Nature}\ }\textbf {\bibinfo {volume} {646}},\ \bibinfo {pages} {1075} (\bibinfo {year} {2025})}\BibitemShut {NoStop}%
\bibitem [{\citenamefont {Acharya}\ \emph {et~al.}(2025)Acharya, Rajeev and Abanin, Dmitry A. and Aghababaie-Beni, Laleh and Aleiner, Igor and Andersen, Trond I. and Ansmann, Markus and Arute, Frank and Arya, Kunal and Asfaw, Abraham and Astrakhantsev, Nikita and Atalaya, Juan and Babbush, Ryan and Bacon, Dave and Ballard, Brian and Bardin, Joseph C. and Bausch, Johannes and Bengtsson, Andreas and Bilmes, Alexander and Blackwell, Sam and Boixo, Sergio and Bortoli, Gina and Bourassa, Alexandre and Bovaird, Jenna and Brill, Leon and Broughton, Michael and Browne, David A. and Buchea, Brett and Buckley, Bob B. and Buell, David A. and Burger, Tim and Burkett, Brian and Bushnell, Nicholas and Cabrera, Anthony and Campero, Juan and Chang, Hung-Shen and Chen, Yu and Chen, Zijun and Chiaro, Ben and Chik, Desmond and Chou, Charina and Claes, Jahan and Cleland, Agnetta Y. and Cogan, Josh and Collins, Roberto and Conner, Paul and Courtney, William and Crook, Alexander L. and Curtin, Ben and Das, Sayan and Davies, Alex
  and De Lorenzo, Laura and Debroy, Dripto M. and Demura, Sean and Devoret, Michel and Di Paolo, Agustin and Donohoe, Paul and Drozdov, Ilya and Dunsworth, Andrew and Earle, Clint and Edlich, Thomas and Eickbusch, Alec and Elbag, Aviv Moshe and Elzouka, Mahmoud and Erickson, Catherine and Faoro, Lara and Farhi, Edward and Ferreira, Vinicius S. and Burgos, Leslie Flores and Forati, Ebrahim and Fowler, Austin G. and Foxen, Brooks and Ganjam, Suhas and Garcia, Gonzalo and Gasca, Robert and Genois, {\'E}lie and Giang, William and Gidney, Craig and Gilboa, Dar and Gosula, Raja and Dau, Alejandro Grajales and Graumann, Dietrich and Greene, Alex and Gross, Jonathan A. and Habegger, Steve and Hall, John and Hamilton, Michael C. and Hansen, Monica and Harrigan, Matthew P. and Harrington, Sean D. and Heras, Francisco J. H. and Heslin, Stephen and Heu, Paula and Higgott, Oscar and Hill, Gordon and Hilton, Jeremy and Holland, George and Hong, Sabrina and Huang, Hsin-Yuan and Huff, Ashley and Huggins, William J. and
  Ioffe, Lev B. and Isakov, Sergei V. and Iveland, Justin and Jeffrey, Evan and Jiang, Zhang and Jones, Cody and Jordan, Stephen and Joshi, Chaitali and Juhas, Pavol and Kafri, Dvir and Kang, Hui and Karamlou, Amir H. and Kechedzhi, Kostyantyn and Kelly, Julian and Khaire, Trupti and Khattar, Tanuj and Khezri, Mostafa and Kim, Seon and Klimov, Paul V. and Klots, Andrey R. and Kobrin, Bryce and Kohli, Pushmeet and Korotkov, Alexander N. and Kostritsa, Fedor and Kothari, Robin and Kozlovskii, Borislav and Kreikebaum, John Mark and Kurilovich, Vladislav D. and Lacroix, Nathan and Landhuis, David and Lange-Dei, Tiano and Langley, Brandon W. and Laptev, Pavel and Lau, Kim-Ming and Le Guevel, Lo{\"i}ck and Ledford, Justin and Lee, Joonho and Lee, Kenny and Lensky, Yuri D. and Leon, Shannon and Lester, Brian J. and Li, Wing Yan and Li, Yin and Lill, Alexander T. and Liu, Wayne and Livingston, William P. and Locharla, Aditya and Lucero, Erik and Lundahl, Daniel and Lunt, Aaron and Madhuk, Sid and Malone, Fionn D. and
  Maloney, Ashley and Mandr{\`a}, Salvatore and Manyika, James and Martin, Leigh S. and Martin, Orion and Martin, Steven and Maxfield, Cameron and McClean, Jarrod R. and McEwen, Matt and Meeks, Seneca and Megrant, Anthony and Mi, Xiao and Miao, Kevin C. and Mieszala, Amanda and Molavi, Reza and Molina, Sebastian and Montazeri, Shirin and Morvan, Alexis and Movassagh, Ramis and Mruczkiewicz, Wojciech and Naaman, Ofer and Neeley, Matthew and Neill, Charles and Nersisyan, Ani and Neven, Hartmut and Newman, Michael and Ng, Jiun How and Nguyen, Anthony and Nguyen, Murray and Ni, Chia-Hung and Niu, Murphy Yuezhen and O'Brien, Thomas E. and Oliver, William D. and Opremcak, Alex and Ottosson, Kristoffer and Petukhov, Andre and Pizzuto, Alex and Platt, John and Potter, Rebecca and Pritchard, Orion and Pryadko, Leonid P. and Quintana, Chris and Ramachandran, Ganesh and Reagor, Matthew J. and Redding, John and Rhodes, David M. and Roberts, Gabrielle and Rosenberg, Eliott and Rosenfeld, Emma and Roushan, Pedram and
  Rubin, Nicholas C. and Saei, Negar and Sank, Daniel and Sankaragomathi, Kannan and Satzinger, Kevin J. and Schurkus, Henry F. and Schuster, Christopher and Senior, Andrew W. and Shearn, Michael J. and Shorter, Aaron and Shutty, Noah and Shvarts, Vladimir and Singh, Shraddha and Sivak, Volodymyr and Skruzny, Jindra and Small, Spencer and Smelyanskiy, Vadim and Smith, W. Clarke and Somma, Rolando D. and Springer, Sofia and Sterling, George and Strain, Doug and Suchard, Jordan and Szasz, Aaron and Sztein, Alex and Thor, Douglas and Torres, Alfredo and Torunbalci, M. Mert and Vaishnav, Abeer and Vargas, Justin and Vdovichev, Sergey and Vidal, Guifre and Villalonga, Benjamin and Heidweiller, Catherine Vollgraff and Waltman, Steven and Wang, Shannon X. and Ware, Brayden and Weber, Kate and Weidel, Travis and White, Theodore and Wong, Kristi and Woo, Bryan W. K. and Xing, Cheng and Yao, Z. Jamie and Yeh, Ping and Ying, Bicheng and Yoo, Juhwan and Yosri, Noureldin and Young, Grayson and Zalcman, Adam and Zhang,
  Yaxing and Zhu, Ningfeng and Zobrist, Nicholas and AI, Google Quantum and {Collaborators}}]{Acharya2025}%
  \BibitemOpen
  \bibfield  {author} {\bibinfo {author} {Acharya, R.}, \bibinfo {author} {Abanin, D.~A.}, \bibinfo {author} {Aghababaie-Beni, L.}, \emph {et~al.},\ }\bibfield  {title} {\bibinfo {title} {Quantum error correction below the surface code threshold},\ }\href {https://doi.org/10.1038/s41586-024-08449-y} {\bibfield  {journal} {\bibinfo  {journal} {Nature}\ }\textbf {\bibinfo {volume} {638}},\ \bibinfo {pages} {920} (\bibinfo {year} {2025})}\BibitemShut {NoStop}%
\bibitem [{\citenamefont {Bluvstein}\ \emph {et~al.}(2023)Bluvstein, Dolev and Evered, Simon J. and Geim, Alexandra A. and Li, Sophie H. and Zhou, Hengyun and Manovitz, Tom and Ebadi, Sepehr and Cain, Madelyn and Kalinowski, Marcin and Hangleiter, Dominik and Ataides, J. Pablo Bonilla and Maskara, Nishad and Cong, Iris and Gao, Xun and Rodriguez, Pedro Sales and Karolyshyn, Thomas and Semeghini, Giulia and Gullans, Michael J. and Greiner, Markus and Vuleti{\'{c}}, Vladan and Lukin, Mikhail D.}]{Bluvstein2023}%
  \BibitemOpen
  \bibfield  {author} {\bibinfo {author} {Bluvstein, D.}, \bibinfo {author} {Evered, S.~J.}, \bibinfo {author} {Geim, A.~A.}, \emph {et~al.},\ }\bibfield  {title} {\bibinfo {title} {Logical quantum processor based on reconfigurable atom arrays},\ }\href {https://doi.org/10.1038/s41586-023-06927-3} {\bibfield  {journal} {\bibinfo  {journal} {Nature}\ }\textbf {\bibinfo {volume} {626}},\ \bibinfo {pages} {58} (\bibinfo {year} {2023})}\BibitemShut {NoStop}%
\bibitem [{\citenamefont {Abanin}\ \emph {et~al.}(2025)Abanin, Dmitry A. and Acharya, Rajeev and Aghababaie-Beni, Laleh and Aigeldinger, Georg and Ajoy, Ashok and Alcaraz, Ross and Aleiner, Igor and Andersen, Trond I. and Ansmann, Markus and Arute, Frank and Arya, Kunal and Asfaw, Abraham and Astrakhantsev, Nikita and Atalaya, Juan and Babbush, Ryan and Bacon, Dave and Ballard, Brian and Bardin, Joseph C. and Bengs, Christian and Bengtsson, Andreas and Bilmes, Alexander and Boixo, Sergio and Bortoli, Gina and Bourassa, Alexandre and Bovaird, Jenna and Bowers, Dylan and Brill, Leon and Broughton, Michael and Browne, David A. and Buchea, Brett and Buckley, Bob B. and Buell, David A. and Burger, Tim and Burkett, Brian and Bushnell, Nicholas and Cabrera, Anthony and Campero, Juan and Chang, Hung-Shen and Chen, Yu and Chen, Zijun and Chiaro, Ben and Chih, Liang-Ying and Chik, Desmond and Chou, Charina and Claes, Jahan and Cleland, Agnetta Y. and Cogan, Josh and Cohen, Saul and Collins, Roberto and Conner, Paul and
  Courtney, William and Crook, Alexander L. and Curtin, Ben and Das, Sayan and De Lorenzo, Laura and Debroy, Dripto M. and Demura, Sean and Devoret, Michel and Di Paolo, Agustin and Donohoe, Paul and Drozdov, Ilya and Dunsworth, Andrew and Earle, Clint and Eickbusch, Alec and Elbag, Aviv Moshe and Elzouka, Mahmoud and Erickson, Catherine and Faoro, Lara and Farhi, Edward and Ferreira, Vinicius S. and Burgos, Leslie Flores and Forati, Ebrahim and Fowler, Austin G. and Foxen, Brooks and Ganjam, Suhas and Garcia, Gonzalo and Gasca, Robert and Genois, Élie and Giang, William and Gidney, Craig and Gilboa, Dar and Gosula, Raja and Dau, Alejandro Grajales and Graumann, Dietrich and Greene, Alex and Gross, Jonathan A. and Gu, Hanfeng and Habegger, Steve and Hall, John and Hamamura, Ikko and Hamilton, Michael C. and Hansen, Monica and Harrigan, Matthew P. and Harrington, Sean D. and Heslin, Stephen and Heu, Paula and Higgott, Oscar and Hill, Gordon and Hilton, Jeremy and Hong, Sabrina and Huang, Hsin-Yuan and Huff,
  Ashley and Huggins, William J. and Ioffe, Lev B. and Isakov, Sergei V. and Iveland, Justin and Jeffrey, Evan and Jiang, Zhang and Jin, Xiaoxuan and Jones, Cody and Jordan, Stephen and Joshi, Chaitali and Juhas, Pavol and Kabel, Andreas and Kafri, Dvir and Kang, Hui and Karamlou, Amir H. and Kechedzhi, Kostyantyn and Kelly, Julian and Khaire, Trupti and Khattar, Tanuj and Khezri, Mostafa and Kim, Seon and King, Robbie and Klimov, Paul V. and Klots, Andrey R. and Kobrin, Bryce and Korotkov, Alexander N. and Kostritsa, Fedor and Kothari, Robin and Kreikebaum, John Mark and Kurilovich, Vladislav D. and Kyoseva, Elica and Landhuis, David and Lange-Dei, Tiano and Langley, Brandon W. and Laptev, Pavel and Lau, Kim-Ming and Le Guevel, Loïck and Ledford, Justin and Lee, Joonho and Lee, Kenny and Lensky, Yuri D. and Leon, Shannon and Lester, Brian J. and Li, Wing Yan and Lill, Alexander T. and Liu, Wayne and Livingston, William P. and Locharla, Aditya and Lucero, Erik and Lundahl, Daniel and Lunt, Aaron and Madhuk,
  Sid and Malone, Fionn D. and Maloney, Ashley and Mandrà, Salvatore and Manyika, James M. and Martin, Leigh S. and Martin, Orion and Martin, Steven and Matias, Yossi and Maxfield, Cameron and McClean, Jarrod R. and McEwen, Matt and Meeks, Seneca and Megrant, Anthony and Mi, Xiao and Miao, Kevin C. and Mieszala, Amanda and Minev, Zlatko and Molavi, Reza and Molina, Sebastian and Montazeri, Shirin and Morvan, Alexis and Movassagh, Ramis and Mruczkiewicz, Wojciech and Naaman, Ofer and Neeley, Matthew and Neill, Charles and Nersisyan, Ani and Neven, Hartmut and Newman, Michael and Ng, Jiun How and Nguyen, Anthony and Nguyen, Murray and Ni, Chia-Hung and Niu, Murphy Yuezhen and Oas, Logan and O’Brien, Thomas E. and Oliver, William D. and Opremcak, Alex and Ottosson, Kristoffer and Petukhov, Andre and Pizzuto, Alex and Platt, John and Potter, Rebecca and Pritchard, Orion and Pryadko, Leonid P. and Quintana, Chris and Ramachandran, Ganesh and Ramanathan, Chandrasekhar and Reagor, Matthew J. and Redding, John and
  Rhodes, David M. and Roberts, Gabrielle and Rosenberg, Eliott and Rosenfeld, Emma and Roushan, Pedram and Rubin, Nicholas C. and Saei, Negar and Sank, Daniel and Sankaragomathi, Kannan and Satzinger, Kevin J. and Schmidhuber, Alexander and Schurkus, Henry F. and Schuster, Christopher and Schuster, Thomas and Shearn, Michael J. and Shorter, Aaron and Shutty, Noah and Shvarts, Vladimir and Sivak, Volodymyr and Skruzny, Jindra and Small, Spencer and Smelyanskiy, Vadim and Smith, W. Clarke and Somma, Rolando D. and Springer, Sofia and Sterling, George and Strain, Doug and Suchard, Jordan and Suchsland, Philippe and Szasz, Aaron and Sztein, Alex and Thor, Douglas and Tomita, Eifu and Torres, Alfredo and Torunbalci, M. Mert and Vaishnav, Abeer and Vargas, Justin and Vdovichev, Sergey and Vidal, Guifre and Villalonga, Benjamin and Heidweiller, Catherine Vollgraff and Waltman, Steven and Wang, Shannon X. and Ware, Brayden and Weber, Kate and Weidel, Travis and Westerhout, Tom and White, Theodore and Wong, Kristi
  and Woo, Bryan W. K. and Xing, Cheng and Yao, Z. Jamie and Yeh, Ping and Ying, Bicheng and Yoo, Juhwan and Yosri, Noureldin and Young, Grayson and Zalcman, Adam and Zhang, Chongwei and Zhang, Yaxing and Zhu, Ningfeng and Zobrist, Nicholas and Google Quantum, A. I. and Collaborators}]{Abanin2025}%
  \BibitemOpen
  \bibfield  {author} {\bibinfo {author} {Abanin, D.~A.}, \bibinfo {author} {Acharya, R.}, \bibinfo {author} {Aghababaie-Beni, L.}, \emph {et~al.},\ }\bibfield  {title} {\bibinfo {title} {Observation of constructive interference at the edge of quantum ergodicity},\ }\href {https://doi.org/10.1038/s41586-025-09526-6} {\bibfield  {journal} {\bibinfo  {journal} {Nature}\ }\textbf {\bibinfo {volume} {646}},\ \bibinfo {pages} {825} (\bibinfo {year} {2025})}\BibitemShut {NoStop}%
\bibitem [{\citenamefont {Devitt}(2016)Devitt, Simon J.}]{Devitt2016}%
  \BibitemOpen
  \bibfield  {author} {\bibinfo {author} {Devitt, S.~J.},\ }\bibfield  {title} {\bibinfo {title} {Performing quantum computing experiments in the cloud},\ }\href {https://doi.org/10.1103/PhysRevA.94.032329} {\bibfield  {journal} {\bibinfo  {journal} {Phys. Rev. A}\ }\textbf {\bibinfo {volume} {94}},\ \bibinfo {pages} {032329} (\bibinfo {year} {2016})}\BibitemShut {NoStop}%
\bibitem [{\citenamefont {Blinov}\ \emph {et~al.}(2021)Blinov, Sergey and Wu, B and Monroe, C}]{Blinov2021}%
  \BibitemOpen
  \bibfield  {author} {\bibinfo {author} {Blinov, S.}, \bibinfo {author} {Wu, B.},\ and\ \bibinfo {author} {Monroe, C.},\ }\bibfield  {title} {\bibinfo {title} {Comparison of cloud-based ion trap and superconducting quantum computer architectures},\ }\href {https://doi.org/10.1116/5.0058187} {\bibfield  {journal} {\bibinfo  {journal} {AVS Quantum Sci.}\ }\textbf {\bibinfo {volume} {3}} (\bibinfo {year} {2021})}\BibitemShut {NoStop}%
\bibitem [{\citenamefont {Wurtz}\ \emph {et~al.}(2023)Jonathan Wurtz and Alexei Bylinskii and Boris Braverman and Jesse Amato-Grill and Sergio H. Cantu and Florian Huber and Alexander Lukin and Fangli Liu and Phillip Weinberg and John Long and Sheng-Tao Wang and Nathan Gemelke and Alexander Keesling}]{Wurtz2023}%
  \BibitemOpen
  \bibfield  {author} {\bibinfo {author} {Wurtz, J.}, \bibinfo {author} {Bylinskii, A.}, \bibinfo {author} {Braverman, B.}, \emph {et~al.},\ }\bibfield  {title} {\bibinfo {title} {{Aquila: QuEra's 256-qubit neutral-atom quantum computer}},\ }\href {https://arxiv.org/abs/2306.11727} {\bibfield  {journal} {\bibinfo  {journal} {arXiv:2306.11727}\ } (\bibinfo {year} {2023})}\BibitemShut {NoStop}%
\bibitem [{\citenamefont {Childs}(2005)Childs, Andrew M.}]{Childs2005b}%
  \BibitemOpen
  \bibfield  {author} {\bibinfo {author} {Childs, A.~M.},\ }\bibfield  {title} {\bibinfo {title} {Secure assisted quantum computation},\ }\href {http://dl.acm.org/citation.cfm?id=2011670.2011674} {\bibfield  {journal} {\bibinfo  {journal} {Quantum Info. Comput.}\ }\textbf {\bibinfo {volume} {5}},\ \bibinfo {pages} {456} (\bibinfo {year} {2005})}\BibitemShut {NoStop}%
\bibitem [{\citenamefont {Arrighi}\ and\ \citenamefont {Salvail}(2006)Arrighi, Pablo and Salvail, Louis}]{Arrighi2006}%
  \BibitemOpen
  \bibfield  {author} {\bibinfo {author} {Arrighi, P.}\ and\ \bibinfo {author} {Salvail, L.},\ }\bibfield  {title} {\bibinfo {title} {Blind quantum computation},\ }\href {https://doi.org/https://doi.org/10.1142/S0219749906002171} {\bibfield  {journal} {\bibinfo  {journal} {Int. J. Quantum Inf.}\ }\textbf {\bibinfo {volume} {4}},\ \bibinfo {pages} {883} (\bibinfo {year} {2006})}\BibitemShut {NoStop}%
\bibitem [{\citenamefont {Raussendorf}\ and\ \citenamefont {Briegel}(2001)Raussendorf, Robert and Briegel, Hans J.}]{Raussendorf2001}%
  \BibitemOpen
  \bibfield  {author} {\bibinfo {author} {Raussendorf, R.}\ and\ \bibinfo {author} {Briegel, H.~J.},\ }\bibfield  {title} {\bibinfo {title} {A one-way quantum computer},\ }\href {https://doi.org/10.1103/PhysRevLett.86.5188} {\bibfield  {journal} {\bibinfo  {journal} {Phys. Rev. Lett.}\ }\textbf {\bibinfo {volume} {86}},\ \bibinfo {pages} {5188} (\bibinfo {year} {2001})}\BibitemShut {NoStop}%
\bibitem [{\citenamefont {Fitzsimons}(2017)Fitzsimons, Joseph F.}]{Fitzsimons2017}%
  \BibitemOpen
  \bibfield  {author} {\bibinfo {author} {Fitzsimons, J.~F.},\ }\bibfield  {title} {\bibinfo {title} {Private quantum computation: an introduction to blind quantum computing and related protocols},\ }\href {https://doi.org/10.1038/s41534-017-0025-3} {\bibfield  {journal} {\bibinfo  {journal} {npj Quantum Inf.}\ }\textbf {\bibinfo {volume} {3}},\ \bibinfo {pages} {23} (\bibinfo {year} {2017})}\BibitemShut {NoStop}%
\bibitem [{\citenamefont {Broadbent}\ \emph {et~al.}(2009)Broadbent, Anne and Fitzsimons, Joseph and Kashefi, Elham}]{Broadbent2009}%
  \BibitemOpen
  \bibfield  {author} {\bibinfo {author} {Broadbent, A.}, \bibinfo {author} {Fitzsimons, J.},\ and\ \bibinfo {author} {Kashefi, E.},\ }\bibfield  {title} {\bibinfo {title} {Universal blind quantum computation},\ }in\ \href {https://doi.org/10.1109/FOCS.2009.36} {\emph {\bibinfo {booktitle} {2009 50th Annual IEEE Symposium on Foundations of Computer Science}}}\ (\bibinfo {year} {2009})\ pp.\ \bibinfo {pages} {517--526}\BibitemShut {NoStop}%
\bibitem [{\citenamefont {Morimae}\ and\ \citenamefont {Fujii}(2013)Morimae, Tomoyuki and Fujii, Keisuke}]{Morimae2013}%
  \BibitemOpen
  \bibfield  {author} {\bibinfo {author} {Morimae, T.}\ and\ \bibinfo {author} {Fujii, K.},\ }\bibfield  {title} {\bibinfo {title} {Blind quantum computation protocol in which alice only makes measurements},\ }\href {https://doi.org/10.1103/PhysRevA.87.050301} {\bibfield  {journal} {\bibinfo  {journal} {Phys. Rev. A}\ }\textbf {\bibinfo {volume} {87}},\ \bibinfo {pages} {050301} (\bibinfo {year} {2013})}\BibitemShut {NoStop}%
\bibitem [{\citenamefont {Barz}\ \emph {et~al.}(2012)Barz, Stefanie and Kashefi, Elham and Broadbent, Anne and Fitzsimons, Joseph F. and Zeilinger, Anton and Walther, Philip}]{Barz2012}%
  \BibitemOpen
  \bibfield  {author} {\bibinfo {author} {Barz, S.}, \bibinfo {author} {Kashefi, E.}, \bibinfo {author} {Broadbent, A.}, \emph {et~al.},\ }\bibfield  {title} {\bibinfo {title} {Demonstration of blind quantum computing},\ }\href {http://www.sciencemag.org/content/335/6066/303.abstract} {\bibfield  {journal} {\bibinfo  {journal} {Science}\ }\textbf {\bibinfo {volume} {335}},\ \bibinfo {pages} {303} (\bibinfo {year} {2012})}\BibitemShut {NoStop}%
\bibitem [{\citenamefont {Barz}\ \emph {et~al.}(2013)Barz, Stefanie and Fitzsimons, Joseph F and Kashefi, Elham and Walther, Philip}]{Barz2013}%
  \BibitemOpen
  \bibfield  {author} {\bibinfo {author} {Barz, S.}, \bibinfo {author} {Fitzsimons, J.~F.}, \bibinfo {author} {Kashefi, E.},\ and\ \bibinfo {author} {Walther, P.},\ }\bibfield  {title} {\bibinfo {title} {Experimental verification of quantum computation},\ }\href {https://doi.org/https://doi.org/10.1038/nphys2763} {\bibfield  {journal} {\bibinfo  {journal} {Nat. Phys.}\ }\textbf {\bibinfo {volume} {9}},\ \bibinfo {pages} {727} (\bibinfo {year} {2013})}\BibitemShut {NoStop}%
\bibitem [{\citenamefont {Fisher}\ \emph {et~al.}(2014)Fisher, Kent AG and Broadbent, Anne and Shalm, LK and Yan, Z and Lavoie, Jonathan and Prevedel, Robert and Jennewein, Thomas and Resch, Kevin J}]{Fisher2014}%
  \BibitemOpen
  \bibfield  {author} {\bibinfo {author} {Fisher, K.~A.}, \bibinfo {author} {Broadbent, A.}, \bibinfo {author} {Shalm, L.}, \emph {et~al.},\ }\bibfield  {title} {\bibinfo {title} {Quantum computing on encrypted data},\ }\href {https://doi.org/10.1038/ncomms4074} {\bibfield  {journal} {\bibinfo  {journal} {Nat. Commun.}\ }\textbf {\bibinfo {volume} {5}},\ \bibinfo {pages} {3074} (\bibinfo {year} {2014})}\BibitemShut {NoStop}%
\bibitem [{\citenamefont {Greganti}\ \emph {et~al.}(2016)Greganti, Chiara and Roehsner, Marie-Christine and Barz, Stefanie and Morimae, Tomoyuki and Walther, Philip}]{Greganti2016}%
  \BibitemOpen
  \bibfield  {author} {\bibinfo {author} {Greganti, C.}, \bibinfo {author} {Roehsner, M.-C.}, \bibinfo {author} {Barz, S.}, \emph {et~al.},\ }\bibfield  {title} {\bibinfo {title} {Demonstration of measurement-only blind quantum computing},\ }\href {https://doi.org/10.1088/1367-2630/18/1/013020} {\bibfield  {journal} {\bibinfo  {journal} {New J. Phys.}\ }\textbf {\bibinfo {volume} {18}},\ \bibinfo {pages} {013020} (\bibinfo {year} {2016})}\BibitemShut {NoStop}%
\bibitem [{\citenamefont {Huang}\ \emph {et~al.}(2017)Huang, He-Liang and Zhao, Qi and Ma, Xiongfeng and Liu, Chang and Su, Zu-En and Wang, Xi-Lin and Li, Li and Liu, Nai-Le and Sanders, Barry C. and Lu, Chao-Yang and Pan, Jian-Wei}]{Huang2017}%
  \BibitemOpen
  \bibfield  {author} {\bibinfo {author} {Huang, H.-L.}, \bibinfo {author} {Zhao, Q.}, \bibinfo {author} {Ma, X.}, \emph {et~al.},\ }\bibfield  {title} {\bibinfo {title} {Experimental blind quantum computing for a classical client},\ }\href {https://doi.org/10.1103/PhysRevLett.119.050503} {\bibfield  {journal} {\bibinfo  {journal} {Phys. Rev. Lett.}\ }\textbf {\bibinfo {volume} {119}},\ \bibinfo {pages} {050503} (\bibinfo {year} {2017})}\BibitemShut {NoStop}%
\bibitem [{\citenamefont {Drmota}\ \emph {et~al.}(2024)Drmota, P. and Nadlinger, D. P. and Main, D. and Nichol, B. C. and Ainley, E. M. and Leichtle, D. and Mantri, A. and Kashefi, E. and Srinivas, R. and Araneda, G. and Ballance, C. J. and Lucas, D. M.}]{Drmota2024}%
  \BibitemOpen
  \bibfield  {author} {\bibinfo {author} {Drmota, P.}, \bibinfo {author} {Nadlinger, D.~P.}, \bibinfo {author} {Main, D.}, \emph {et~al.},\ }\bibfield  {title} {\bibinfo {title} {Verifiable blind quantum computing with trapped ions and single photons},\ }\href {https://doi.org/10.1103/PhysRevLett.132.150604} {\bibfield  {journal} {\bibinfo  {journal} {Phys. Rev. Lett.}\ }\textbf {\bibinfo {volume} {132}},\ \bibinfo {pages} {150604} (\bibinfo {year} {2024})}\BibitemShut {NoStop}%
\bibitem [{\citenamefont {Wei}\ \emph {et~al.}(2025)Y.-C. Wei and P.-J. Stas and A. Suleymanzade and G. Baranes and F. Machado and Y. Q. Huan and C. M. Knaut and S. W. Ding and M. Merz and E. N. Knall and U. Yazlar and M. Sirotin and I. W. Wang and B. Machielse and S. F. Yelin and J. Borregaard and H. Park and M. Lončar and M. D. Lukin}]{Wei2025}%
  \BibitemOpen
  \bibfield  {author} {\bibinfo {author} {Wei, Y.-C.}, \bibinfo {author} {Stas, P.-J.}, \bibinfo {author} {Suleymanzade, A.}, \emph {et~al.},\ }\bibfield  {title} {\bibinfo {title} {Universal distributed blind quantum computing with solid-state qubits},\ }\href {https://doi.org/10.1126/science.adu6894} {\bibfield  {journal} {\bibinfo  {journal} {Science}\ }\textbf {\bibinfo {volume} {388}},\ \bibinfo {pages} {509} (\bibinfo {year} {2025})}\BibitemShut {NoStop}%
\bibitem [{\citenamefont {Pathumsoot}\ \emph {et~al.}(2020)Pathumsoot, Poramet and Matsuo, Takaaki and Satoh, Takahiko and Hajdu{\v{s}}ek, Michal and Suwanna, Sujin and Van Meter, Rodney}]{Pathumsoot2020}%
  \BibitemOpen
  \bibfield  {author} {\bibinfo {author} {Pathumsoot, P.}, \bibinfo {author} {Matsuo, T.}, \bibinfo {author} {Satoh, T.}, \emph {et~al.},\ }\bibfield  {title} {\bibinfo {title} {Modeling of measurement-based quantum network coding on a superconducting quantum processor},\ }\href {https://doi.org/10.1103/PhysRevA.101.052301} {\bibfield  {journal} {\bibinfo  {journal} {Phys. Rev. A}\ }\textbf {\bibinfo {volume} {101}},\ \bibinfo {pages} {052301} (\bibinfo {year} {2020})}\BibitemShut {NoStop}%
\bibitem [{\citenamefont {Jiang}\ \emph {et~al.}(2026)Jiang, Tao and Cai, Jianbin and Huang, Junxiang and Zhou, Naibin and Zhang, Yukun and Bei, Jiahao and Cai, Guoqing and Cao, Sirui and Chen, Fusheng and Chen, Jiang and Chen, Kefu and Chen, Xiawei and Chen, Xiqing and Chen, Zhe and Chen, Zhiyuan and Chen, Zihua and Chu, Wenhao and Deng, Hui and Deng, Zhibin and Ding, Pei and Ding, Xun and Ding, Zhuzhengqi and Dong, Shuai and Fan, Bo and Fan, Daojin and Fu, Yuanhao and Gao, Dongxin and Ge, Lei and Gui, Jiacheng and Guo, Cheng and Guo, Shaojun and Guo, Xiaoyang and Han, Lianchen and He, Tan and Hong, Linyin and Hu, Yisen and Huang, He-Liang and Huo, Yong-Heng and Jiang, Zuokai and Jin, Honghong and Leng, Yunxiang and Li, Dayu and Li, Dongdong and Li, Fangyu and Li, Jiaqi and Li, Jinjin and Li, Junyan and Li, Junyun and Li, Na and Li, Shaowei and Li, Wei and Li, Yuhuai and Li, Yuan and Liang, Futian and Liang, Xuelian and Liao, Nanxing and Lin, Jin and Lin, Weiping and Liu, Dailin and Liu, Hongxiu and Liu,
  Maliang and Liu, Xinyu and Liu, Xuemeng and Liu, Yancheng and Lou, Haoxin and Ma, Yuwei and Meng, Lingxin and Mou, Hao and Nan, Kailiang and Nie, Binghan and Nie, Meijuan and Ning, Jie and Niu, Le and Peng, Wenyi and Qian, Haoran and Rong, Hao and Rong, Tao and Shen, Huiyan and Shen, Qiong and Su, Hong and Su, Feifan and Sun, Chenyin and Sun, Liangchao and Sun, Tianzuo and Sun, Yingxiu and Tan, Yimeng and Tan, Jun and Tang, Longyue and Tu, Wenbing and Wang, Jiafei and Wang, Biao and Wang, Chang and Wang, Chen and Wang, Chu and Wang, Jian and Wang, Liangyuan and Wang, Rui and Wang, Shengtao and Wang, Xiaomin and Wang, Xinzhe and Wang, Xunxun and Wang, Yeru and Wei, Zuolin and Wei, Jiazhou and Wu, Dachao and Wu, Gang and Wu, Jin and Wu, Yulin and Xie, Shiyong and Xin, Lianjie and Xu, Yu and Xue, Chun and Yan, Kai and Yang, Weifeng and Yang, Xinpeng and Yang, Yang and Ye, Yangsen and Ye, Zhenping and Ying, Chong and Yu, Jiale and Yu, Qinjing and Yu, Wenhu and Zeng, Xiangdong and Zha, Chen and Zhan, Shaoyu and
  Zhang, Feifei and Zhang, Haibin and Zhang, Kaili and Zhang, Wen and Zhang, Yiming and Zhang, Yongzhuo and Zhang, Lixiang and Zhao, Guming and Zhao, Peng and Zhao, Xintao and Zhao, Youwei and Zhao, Zhong and Zheng, Luyuan and Zhou, Fei and Zhou, Liang and Zhou, Na and Zhou, Shifeng and Zhou, Shuang and Zhou, Zhengxiao and Zhu, Chengjun and Zhu, Qingling and Zou, Guihong and Zou, Haonan and Zhang, Qiang and Lu, Chao-Yang and Peng, Cheng-Zhi and Yuan, Xiao and Gong, Ming and Zhu, Xiaobo and Pan, Jian-Wei}]{Jiang2026}%
  \BibitemOpen
  \bibfield  {author} {\bibinfo {author} {Jiang, T.}, \bibinfo {author} {Cai, J.}, \bibinfo {author} {Huang, J.}, \emph {et~al.},\ }\bibfield  {title} {\bibinfo {title} {One- and two-dimensional cluster states for topological phase simulation and measurement-based quantum computation},\ }\href {https://doi.org/10.1038/s41567-026-03179-6} {\bibfield  {journal} {\bibinfo  {journal} {Nat. Phys.}\ }\textbf {\bibinfo {volume} {22}},\ \bibinfo {pages} {430} (\bibinfo {year} {2026})}\BibitemShut {NoStop}%
\bibitem [{\citenamefont {Yang}\ \emph {et~al.}(2022)Yang, Zhi-Peng and Ku, Huan-Yu and Baishya, Alakesh and Zhang, Yu-Ran and Kockum, Anton Frisk and Chen, Yueh-Nan and Li, Fu-Li and Tsai, Jaw-Shen and Nori, Franco}]{Yang2022b}%
  \BibitemOpen
  \bibfield  {author} {\bibinfo {author} {Yang, Z.-P.}, \bibinfo {author} {Ku, H.-Y.}, \bibinfo {author} {Baishya, A.}, \emph {et~al.},\ }\bibfield  {title} {\bibinfo {title} {Deterministic one-way logic gates on a cloud quantum computer},\ }\href {https://doi.org/10.1103/PhysRevA.105.042610} {\bibfield  {journal} {\bibinfo  {journal} {Phys. Rev. A}\ }\textbf {\bibinfo {volume} {105}},\ \bibinfo {pages} {042610} (\bibinfo {year} {2022})}\BibitemShut {NoStop}%
\bibitem [{\citenamefont {Koch}\ \emph {et~al.}(2007)Koch, J. and Yu, T. M. and Gambetta, J. and Houck, A. A. and Schuster,D. I. and Majer, J. and Blais, A. and Devoret, M. H. and Girvin, S. M. and Schoelkopf, R. J.}]{Koch2007}%
  \BibitemOpen
  \bibfield  {author} {\bibinfo {author} {Koch, J.}, \bibinfo {author} {Yu, T.~M.}, \bibinfo {author} {Gambetta, J.}, \emph {et~al.},\ }\bibfield  {title} {\bibinfo {title} {Charge-insensitive qubit design derived from the {Cooper} pair box},\ }\href {https://doi.org/10.1103/PhysRevA.76.042319} {\bibfield  {journal} {\bibinfo  {journal} {Phys. Rev. A}\ }\textbf {\bibinfo {volume} {76}},\ \bibinfo {eid} {042319} (\bibinfo {year} {2007})}\BibitemShut {NoStop}%
\bibitem [{\citenamefont {Dalton}\ \emph {et~al.}(2025)Dalton, Kieran and Kn\"orzer, Johannes and Hoehne, Finn and Song, Yongxin and Flasby, Alexander and Colao Zanuz, Dante and Bahrami Panah, Mohsen and Besedin, Ilya and Besse, Jean-Claude and Wallraff, Andreas}]{Dalton2025}%
  \BibitemOpen
  \bibfield  {author} {\bibinfo {author} {Dalton, K.}, \bibinfo {author} {Kn\"orzer, J.}, \bibinfo {author} {Hoehne, F.}, \emph {et~al.},\ }\bibfield  {title} {\bibinfo {title} {Resource-efficient cross-platform verification with modular superconducting devices},\ }\href {https://doi.org/10.1103/czph-xpzs} {\bibfield  {journal} {\bibinfo  {journal} {PRX Quantum}\ }\textbf {\bibinfo {volume} {6}},\ \bibinfo {pages} {040365} (\bibinfo {year} {2025})}\BibitemShut {NoStop}%
\bibitem [{\citenamefont {Norris}\ \emph {et~al.}(2026)Norris, Graham J. and Dalton, Kieran and Colao Zanuz, Dante and Rommens, Alexander and Flasby, Alexander and Bahrami Panah, Mohsen and Swiadek, Fran{\c{c}}ois and Scarato, Colin and Hellings, Christoph and Besse, Jean-Claude and Wallraff, Andreas}]{Norris2026}%
  \BibitemOpen
  \bibfield  {author} {\bibinfo {author} {Norris, G.~J.}, \bibinfo {author} {Dalton, K.}, \bibinfo {author} {Colao Zanuz, D.}, \emph {et~al.},\ }\bibfield  {title} {\bibinfo {title} {Performance characterization of a multi-module quantum processor with static inter-chip couplers},\ }\href {https://doi.org/10.1140/epjqt/s40507-026-00469-z} {\bibfield  {journal} {\bibinfo  {journal} {EPJ Quantum Technol.}\ }\textbf {\bibinfo {volume} {13}},\ \bibinfo {pages} {29} (\bibinfo {year} {2026})}\BibitemShut {NoStop}%
\bibitem [{\citenamefont {Raussendorf}\ \emph {et~al.}(2003)Raussendorf, R. and Browne, D.~E. and Briegel, H.~J.}]{Raussendorf2003}%
  \BibitemOpen
  \bibfield  {author} {\bibinfo {author} {Raussendorf, R.}, \bibinfo {author} {Browne, D.~E.},\ and\ \bibinfo {author} {Briegel, H.~J.},\ }\bibfield  {title} {\bibinfo {title} {Measurement-based quantum computation on cluster states},\ }\href {https://doi.org/10.1103/PhysRevA.68.022312} {\bibfield  {journal} {\bibinfo  {journal} {Phys. Rev. A}\ }\textbf {\bibinfo {volume} {68}},\ \bibinfo {eid} {022312} (\bibinfo {year} {2003})}\BibitemShut {NoStop}%
\bibitem [{\citenamefont {Nielsen}(2004)Nielsen, Michael A}]{Nielsen2004}%
  \BibitemOpen
  \bibfield  {author} {\bibinfo {author} {Nielsen, M.~A.},\ }\bibfield  {title} {\bibinfo {title} {Optical quantum computation using cluster states},\ }\href {https://doi.org/https://doi.org/10.1103/PhysRevLett.93.040503} {\bibfield  {journal} {\bibinfo  {journal} {Phys. Rev. Lett.}\ }\textbf {\bibinfo {volume} {93}},\ \bibinfo {pages} {040503} (\bibinfo {year} {2004})}\BibitemShut {NoStop}%
\bibitem [{\citenamefont {Nielsen}(2006)Nielsen, Michael A}]{Nielsen2006}%
  \BibitemOpen
  \bibfield  {author} {\bibinfo {author} {Nielsen, M.~A.},\ }\bibfield  {title} {\bibinfo {title} {Cluster-state quantum computation},\ }\href {https://doi.org/https://doi.org/10.1016/S0034-4877(06)80014-5} {\bibfield  {journal} {\bibinfo  {journal} {Rep. Math. Phys.}\ }\textbf {\bibinfo {volume} {57}},\ \bibinfo {pages} {147} (\bibinfo {year} {2006})}\BibitemShut {NoStop}%
\bibitem [{\citenamefont {Flammia}\ and\ \citenamefont {Liu}(2011)Flammia, S.~T. and Liu, Y.-K.}]{Flammia2011}%
  \BibitemOpen
  \bibfield  {author} {\bibinfo {author} {Flammia, S.~T.}\ and\ \bibinfo {author} {Liu, Y.-K.},\ }\bibfield  {title} {\bibinfo {title} {Direct fidelity estimation from few pauli measurements},\ }\href {https://doi.org/10.1103/PhysRevLett.106.230501} {\bibfield  {journal} {\bibinfo  {journal} {Phys. Rev. Lett.}\ }\textbf {\bibinfo {volume} {106}},\ \bibinfo {pages} {230501} (\bibinfo {year} {2011})}\BibitemShut {NoStop}%
\bibitem [{\citenamefont {T\'oth}\ and\ \citenamefont {G\"uhne}(2005)T\'oth, G\'eza and G\"uhne, Otfried}]{Toth2005}%
  \BibitemOpen
  \bibfield  {author} {\bibinfo {author} {T\'oth, G.}\ and\ \bibinfo {author} {G\"uhne, O.},\ }\bibfield  {title} {\bibinfo {title} {Entanglement detection in the stabilizer formalism},\ }\href {https://doi.org/10.1103/PhysRevA.72.022340} {\bibfield  {journal} {\bibinfo  {journal} {Phys. Rev. A}\ }\textbf {\bibinfo {volume} {72}},\ \bibinfo {pages} {022340} (\bibinfo {year} {2005})}\BibitemShut {NoStop}%
\bibitem [{\citenamefont {Chuang}\ and\ \citenamefont {Nielsen}(1997)Chuang, I. L. and Nielsen, M. A.}]{Chuang1997}%
  \BibitemOpen
  \bibfield  {author} {\bibinfo {author} {Chuang, I.~L.}\ and\ \bibinfo {author} {Nielsen, M.~A.},\ }\bibfield  {title} {\bibinfo {title} {Prescription for experimental determination of the dynamics of a quantum black box},\ }\href {https://doi.org/10.1080/09500349708231894} {\bibfield  {journal} {\bibinfo  {journal} {J. Mod. Opt.}\ }\textbf {\bibinfo {volume} {44}},\ \bibinfo {pages} {2455} (\bibinfo {year} {1997})}\BibitemShut {NoStop}%
\bibitem [{\citenamefont {James}\ \emph {et~al.}(2001)D. F. V. James and P. G. Kwiat and W. J. Munro and A. G. White}]{James2001}%
  \BibitemOpen
  \bibfield  {author} {\bibinfo {author} {James, D. F.~V.}, \bibinfo {author} {Kwiat, P.~G.}, \bibinfo {author} {Munro, W.~J.},\ and\ \bibinfo {author} {White, A.~G.},\ }\bibfield  {title} {\bibinfo {title} {Measurement of qubits},\ }\href {https://doi.org/10.1103/PhysRevA.64.052312} {\bibfield  {journal} {\bibinfo  {journal} {Phys. Rev. A}\ }\textbf {\bibinfo {volume} {64}},\ \bibinfo {pages} {052312} (\bibinfo {year} {2001})}\BibitemShut {NoStop}%
\bibitem [{\citenamefont {Vallone}\ \emph {et~al.}(2010)Vallone, Giuseppe and Donati, Gaia and Bruno, Natalia and Chiuri, Andrea and Mataloni, Paolo}]{Vallone2010}%
  \BibitemOpen
  \bibfield  {author} {\bibinfo {author} {Vallone, G.}, \bibinfo {author} {Donati, G.}, \bibinfo {author} {Bruno, N.}, \emph {et~al.},\ }\bibfield  {title} {\bibinfo {title} {{Experimental realization of the Deutsch-Jozsa algorithm with a six-qubit cluster state}},\ }\href {https://doi.org/10.1103/PhysRevA.81.050302} {\bibfield  {journal} {\bibinfo  {journal} {Phys. Rev. A}\ }\textbf {\bibinfo {volume} {81}},\ \bibinfo {pages} {050302} (\bibinfo {year} {2010})}\BibitemShut {NoStop}%
\bibitem [{\citenamefont {Tame}\ and\ \citenamefont {Kim}(2010)Tame, M. S. and Kim, M. S.}]{Tame2010}%
  \BibitemOpen
  \bibfield  {author} {\bibinfo {author} {Tame, M.~S.}\ and\ \bibinfo {author} {Kim, M.~S.},\ }\bibfield  {title} {\bibinfo {title} {{Scalable method for demonstrating the Deutsch-Jozsa and Bernstein-Vazirani algorithms using cluster states}},\ }\href {https://doi.org/10.1103/PhysRevA.82.030305} {\bibfield  {journal} {\bibinfo  {journal} {Phys. Rev. A}\ }\textbf {\bibinfo {volume} {82}},\ \bibinfo {pages} {030305} (\bibinfo {year} {2010})}\BibitemShut {NoStop}%
\bibitem [{\citenamefont {DiCarlo}\ \emph {et~al.}(2009)DiCarlo, L. and Chow, J. M. and Gambetta, J. M. and Bishop, Lev S. and Johnson, B. R. and Schuster, D. I. and Majer, J. and Blais, A. and Frunzio, L. and Girvin, S. M. and Schoelkopf, R. J.}]{DiCarlo2009}%
  \BibitemOpen
  \bibfield  {author} {\bibinfo {author} {DiCarlo, L.}, \bibinfo {author} {Chow, J.~M.}, \bibinfo {author} {Gambetta, J.~M.}, \emph {et~al.},\ }\bibfield  {title} {\bibinfo {title} {Demonstration of two-qubit algorithms with a superconducting quantum processor},\ }\href {https://doi.org/10.1038/nature08121} {\bibfield  {journal} {\bibinfo  {journal} {Nature}\ }\textbf {\bibinfo {volume} {460}},\ \bibinfo {pages} {240} (\bibinfo {year} {2009})}\BibitemShut {NoStop}%
\bibitem [{\citenamefont {Holevo}(1973)Holevo, Alexander Semenovich}]{Holevo1973}%
  \BibitemOpen
  \bibfield  {author} {\bibinfo {author} {Holevo, A.~S.},\ }\bibfield  {title} {\bibinfo {title} {Bounds for the quantity of information transmitted by a quantum communication channel},\ }\href {https://www.mathnet.ru/php/archive.phtml?wshow=paper&jrnid=ppi&paperid=903&option_lang=eng} {\bibfield  {journal} {\bibinfo  {journal} {Problemy Peredachi Informatsii}\ }\textbf {\bibinfo {volume} {9}},\ \bibinfo {pages} {3} (\bibinfo {year} {1973})}\BibitemShut {NoStop}%
\bibitem [{\citenamefont {Raussendorf}\ \emph {et~al.}(2007)R Raussendorf and J Harrington and K Goyal}]{Raussendorf2007a}%
  \BibitemOpen
  \bibfield  {author} {\bibinfo {author} {Raussendorf, R.}, \bibinfo {author} {Harrington, J.},\ and\ \bibinfo {author} {Goyal, K.},\ }\bibfield  {title} {\bibinfo {title} {Topological fault-tolerance in cluster state quantum computation},\ }\href {http://stacks.iop.org/1367-2630/9/i=6/a=199} {\bibfield  {journal} {\bibinfo  {journal} {New J. Phys.}\ }\textbf {\bibinfo {volume} {9}},\ \bibinfo {pages} {199} (\bibinfo {year} {2007})}\BibitemShut {NoStop}%
\bibitem [{\citenamefont {Ramette}\ \emph {et~al.}(2024)Ramette, Joshua and Sinclair, Josiah and Breuckmann, Nikolas P. and Vuleti{\ifmmode\acute{c}\else\'{c}\fi}, Vladan}]{Ramette2023}%
  \BibitemOpen
  \bibfield  {author} {\bibinfo {author} {Ramette, J.}, \bibinfo {author} {Sinclair, J.}, \bibinfo {author} {Breuckmann, N.~P.},\ and\ \bibinfo {author} {Vuleti{\ifmmode\acute{c}\else\'{c}\fi}, V.},\ }\bibfield  {title} {\bibinfo {title} {{Fault-tolerant connection of error-corrected qubits with noisy links}},\ }\href {https://doi.org/10.1038/s41534-024-00855-4} {\bibfield  {journal} {\bibinfo  {journal} {npj Quantum Inf.}\ }\textbf {\bibinfo {volume} {10}},\ \bibinfo {pages} {1} (\bibinfo {year} {2024})}\BibitemShut {NoStop}%
\bibitem [{\citenamefont {Magnard}\ \emph {et~al.}(2020)Magnard, P. and Storz, S. and Kurpiers, P. and Sch\"ar, J. and Marxer, F. and L\"utolf, J. and Walter, T. and Besse, J.-C. and Gabureac, M. and Reuer, K. and Akin, A. and Royer, B. and Blais, A. and Wallraff, A.}]{Magnard2020}%
  \BibitemOpen
  \bibfield  {author} {\bibinfo {author} {Magnard, P.}, \bibinfo {author} {Storz, S.}, \bibinfo {author} {Kurpiers, P.}, \emph {et~al.},\ }\bibfield  {title} {\bibinfo {title} {Microwave quantum link between superconducting circuits housed in spatially separated cryogenic systems},\ }\href {https://doi.org/10.1103/PhysRevLett.125.260502} {\bibfield  {journal} {\bibinfo  {journal} {Phys. Rev. Lett.}\ }\textbf {\bibinfo {volume} {125}},\ \bibinfo {pages} {260502} (\bibinfo {year} {2020})}\BibitemShut {NoStop}%
\bibitem [{\citenamefont {Yam}\ \emph {et~al.}(2025)Yam, W. K. and Renger, M. and Gandorfer, S. and Fesquet, F. and Handschuh, M. and Honasoge, K. E. and Kronowetter, F. and Nojiri, Y. and Partanen, M. and Pfeiffer, M. and van der Vliet, H. and Matthews, A. J. and Govenius, J. and Jabdaraghi, R. N. and Prunnila, M. and Marx, A. and Deppe, F. and Gross, R. and Fedorov, K. G.}]{Yam2025}%
  \BibitemOpen
  \bibfield  {author} {\bibinfo {author} {Yam, W.~K.}, \bibinfo {author} {Renger, M.}, \bibinfo {author} {Gandorfer, S.}, \emph {et~al.},\ }\bibfield  {title} {\bibinfo {title} {Cryogenic microwave link for quantum local area networks},\ }\href {https://doi.org/10.1038/s41534-025-01046-5} {\bibfield  {journal} {\bibinfo  {journal} {npj Quantum Inf.}\ }\textbf {\bibinfo {volume} {11}},\ \bibinfo {pages} {87} (\bibinfo {year} {2025})}\BibitemShut {NoStop}%
\bibitem [{\citenamefont {Han}\ \emph {et~al.}(2021)Xu Han and Wei Fu and Chang-Ling Zou and Liang Jiang and Hong X. Tang}]{Han2021f}%
  \BibitemOpen
  \bibfield  {author} {\bibinfo {author} {Han, X.}, \bibinfo {author} {Fu, W.}, \bibinfo {author} {Zou, C.-L.}, \emph {et~al.},\ }\bibfield  {title} {\bibinfo {title} {Microwave-optical quantum frequency conversion},\ }\href {https://doi.org/10.1364/OPTICA.425414} {\bibfield  {journal} {\bibinfo  {journal} {Optica}\ }\textbf {\bibinfo {volume} {8}},\ \bibinfo {pages} {1050} (\bibinfo {year} {2021})}\BibitemShut {NoStop}%
\bibitem [{\citenamefont {Schwartz}\ \emph {et~al.}(2016)Schwartz, I. and Cogan, D. and Schmidgall, E. R. and Don, Y. and Gantz, L. and Kenneth, O. and Lindner, N. H. and Gershoni, D.}]{Schwartz2016}%
  \BibitemOpen
  \bibfield  {author} {\bibinfo {author} {Schwartz, I.}, \bibinfo {author} {Cogan, D.}, \bibinfo {author} {Schmidgall, E.~R.}, \emph {et~al.},\ }\bibfield  {title} {\bibinfo {title} {Deterministic generation of a cluster state of entangled photons},\ }\href {https://doi.org/10.1126/science.aah4758} {\bibfield  {journal} {\bibinfo  {journal} {Science}\ }\textbf {\bibinfo {volume} {354}},\ \bibinfo {pages} {434} (\bibinfo {year} {2016})}\BibitemShut {NoStop}%
\bibitem [{\citenamefont {Larsen}\ \emph {et~al.}(2019)Larsen, Mikkel V. and Guo, Xueshi and Breum, Casper R. and Neergaard-Nielsen, Jonas S. and Andersen, Ulrik L.}]{Larsen2019}%
  \BibitemOpen
  \bibfield  {author} {\bibinfo {author} {Larsen, M.~V.}, \bibinfo {author} {Guo, X.}, \bibinfo {author} {Breum, C.~R.}, \emph {et~al.},\ }\bibfield  {title} {\bibinfo {title} {Deterministic generation of a two-dimensional cluster state},\ }\href {https://doi.org/10.1126/science.aay4354} {\bibfield  {journal} {\bibinfo  {journal} {Science}\ }\textbf {\bibinfo {volume} {366}},\ \bibinfo {pages} {369} (\bibinfo {year} {2019})}\BibitemShut {NoStop}%
\bibitem [{\citenamefont {Ferreira}\ \emph {et~al.}(2024)Ferreira, Vinicius S. and Kim, Gihwan and Butler, Andreas and Pichler, Hannes and Painter, Oskar}]{Ferreira2024}%
  \BibitemOpen
  \bibfield  {author} {\bibinfo {author} {Ferreira, V.~S.}, \bibinfo {author} {Kim, G.}, \bibinfo {author} {Butler, A.}, \emph {et~al.},\ }\bibfield  {title} {\bibinfo {title} {Deterministic generation of multidimensional photonic cluster states with a single quantum emitter},\ }\href {https://doi.org/10.1038/s41567-024-02408-0} {\bibfield  {journal} {\bibinfo  {journal} {Nat. Phys.}\ }\textbf {\bibinfo {volume} {20}},\ \bibinfo {pages} {865} (\bibinfo {year} {2024})}\BibitemShut {NoStop}%
\bibitem [{\citenamefont {O'Sullivan}\ \emph {et~al.}(2025)O'Sullivan, James and Reuer, Kevin and Grigorev, Aleksandr and Dai, Xi and Hernandez-Anton, Alonso and Munoz-Arias, Manuel H. and Hellings, Christoph and Flasby, Alexander and Colao Zanuz, Dante and Besse, Jean-Claude and Blais, Alexandre and Malz, Daniel and Eichler, Christopher and Wallraff, Andreas}]{OSullivan2025}%
  \BibitemOpen
  \bibfield  {author} {\bibinfo {author} {O'Sullivan, J.}, \bibinfo {author} {Reuer, K.}, \bibinfo {author} {Grigorev, A.}, \emph {et~al.},\ }\bibfield  {title} {\bibinfo {title} {Deterministic generation of two-dimensional multi-photon cluster states},\ }\href {https://doi.org/https://doi.org/10.1038/s41467-025-60472-3} {\bibfield  {journal} {\bibinfo  {journal} {Nat. Commun.}\ }\textbf {\bibinfo {volume} {16}},\ \bibinfo {pages} {5505} (\bibinfo {year} {2025})}\BibitemShut {NoStop}%
\bibitem [{\citenamefont {Cao}\ \emph {et~al.}(2023)Cao, Sirui and Wu, Bujiao and Chen, Fusheng and Gong, Ming and Wu, Yulin and Ye, Yangsen and Zha, Chen and Qian, Haoran and Ying, Chong and Guo, Shaojun and Zhu, Qingling and Huang, He-Liang and Zhao, Youwei and Li, Shaowei and Wang, Shiyu and Yu, Jiale and Fan, Daojin and Wu, Dachao and Su, Hong and Deng, Hui and Rong, Hao and Li, Yuan and Zhang, Kaili and Chung, Tung-Hsun and Liang, Futian and Lin, Jin and Xu, Yu and Sun, Lihua and Guo, Cheng and Li, Na and Huo, Yong-Heng and Peng, Cheng-Zhi and Lu, Chao-Yang and Yuan, Xiao and Zhu, Xiaobo and Pan, Jian-Wei}]{Cao2023a}%
  \BibitemOpen
  \bibfield  {author} {\bibinfo {author} {Cao, S.}, \bibinfo {author} {Wu, B.}, \bibinfo {author} {Chen, F.}, \emph {et~al.},\ }\bibfield  {title} {\bibinfo {title} {Generation of genuine entanglement up to 51 superconducting qubits},\ }\href {https://doi.org/10.1038/s41586-023-06195-1} {\bibfield  {journal} {\bibinfo  {journal} {Nature}\ }\textbf {\bibinfo {volume} {619}},\ \bibinfo {pages} {738} (\bibinfo {year} {2023})}\BibitemShut {NoStop}%
\bibitem [{\citenamefont {Schreier}\ \emph {et~al.}(2008)Schreier, J.~A. and Houck, A.~A. and Koch, J. and Schuster, D. I. and Johnson, B.~R. and Chow, J.~M. and Gambetta, J.~M. and Majer, J. and Frunzio, L. and Devoret, M.~H. and Girvin, S.~M. and Schoelkopf, R.~J.}]{Schreier2008}%
  \BibitemOpen
  \bibfield  {author} {\bibinfo {author} {Schreier, J.~A.}, \bibinfo {author} {Houck, A.~A.}, \bibinfo {author} {Koch, J.}, \emph {et~al.},\ }\bibfield  {title} {\bibinfo {title} {Suppressing charge noise decoherence in superconducting charge qubits},\ }\href {https://doi.org/10.1103/PhysRevB.77.180502} {\bibfield  {journal} {\bibinfo  {journal} {Phys. Rev. B}\ }\textbf {\bibinfo {volume} {77}},\ \bibinfo {pages} {180502} (\bibinfo {year} {2008})}\BibitemShut {NoStop}%
\bibitem [{\citenamefont {Motzoi}\ \emph {et~al.}(2009)F. Motzoi and J. M. Gambetta and P. Rebentrost and F. K. Wilhelm}]{Motzoi2009}%
  \BibitemOpen
  \bibfield  {author} {\bibinfo {author} {Motzoi, F.}, \bibinfo {author} {Gambetta, J.~M.}, \bibinfo {author} {Rebentrost, P.},\ and\ \bibinfo {author} {Wilhelm, F.~K.},\ }\bibfield  {title} {\bibinfo {title} {Simple pulses for elimination of leakage in weakly nonlinear qubits},\ }\href {https://doi.org/10.1103/PhysRevLett.103.110501} {\bibfield  {journal} {\bibinfo  {journal} {Phys. Rev. Lett.}\ }\textbf {\bibinfo {volume} {103}},\ \bibinfo {eid} {110501} (\bibinfo {year} {2009})}\BibitemShut {NoStop}%
\bibitem [{\citenamefont {Rol}\ \emph {et~al.}(2019)Rol, M. A. and Battistel, F. and Malinowski, F. K. and Bultink, C. C. and Tarasinski, B. M. and Vollmer, R. and Haider, N. and Muthusubramanian, N. and Bruno, A. and Terhal, B. M. and DiCarlo, L.}]{Rol2019}%
  \BibitemOpen
  \bibfield  {author} {\bibinfo {author} {Rol, M.~A.}, \bibinfo {author} {Battistel, F.}, \bibinfo {author} {Malinowski, F.~K.}, \emph {et~al.},\ }\bibfield  {title} {\bibinfo {title} {Fast, high-fidelity conditional-phase gate exploiting leakage interference in weakly anharmonic superconducting qubits},\ }\href {https://doi.org/10.1103/PhysRevLett.123.120502} {\bibfield  {journal} {\bibinfo  {journal} {Phys. Rev. Lett.}\ }\textbf {\bibinfo {volume} {123}},\ \bibinfo {pages} {120502} (\bibinfo {year} {2019})}\BibitemShut {NoStop}%
\bibitem [{\citenamefont {Negirneac}\ \emph {et~al.}(2021)Negirneac, V. and Ali, H. and Muthusubramanian, N. and Battistel, F. and Sagastizabal, R. and Moreira, M. S. and Marques, J. F. and Vlothuizen, W. J. and Beekman, M. and Zachariadis, C. and Haider, N. and Bruno, A. and DiCarlo, L.}]{Negirneac2021}%
  \BibitemOpen
  \bibfield  {author} {\bibinfo {author} {Negirneac, V.}, \bibinfo {author} {Ali, H.}, \bibinfo {author} {Muthusubramanian, N.}, \emph {et~al.},\ }\bibfield  {title} {\bibinfo {title} {High-fidelity controlled-${Z}$ gate with maximal intermediate leakage operating at the speed limit in a superconducting quantum processor},\ }\href {https://doi.org/10.1103/PhysRevLett.126.220502} {\bibfield  {journal} {\bibinfo  {journal} {Phys. Rev. Lett.}\ }\textbf {\bibinfo {volume} {126}},\ \bibinfo {pages} {220502} (\bibinfo {year} {2021})}\BibitemShut {NoStop}%
\bibitem [{\citenamefont {Hellings}\ \emph {et~al.}(2025)Hellings, Christoph and Lacroix, Nathan and Remm, Ants and Boell, Richard and Herrmann, Johannes and Laz\u{a}r, Stefania and Krinner, Sebastian and Swiadek, Fran\c{c}ois and Andersen, Christian Kraglund and Eichler, Christopher and Wallraff, Andreas}]{Hellings2025}%
  \BibitemOpen
  \bibfield  {author} {\bibinfo {author} {Hellings, C.}, \bibinfo {author} {Lacroix, N.}, \bibinfo {author} {Remm, A.}, \emph {et~al.},\ }\bibfield  {title} {\bibinfo {title} {Calibrating magnetic flux control in superconducting circuits by compensating distortions on timescales from nanoseconds up to tens of microseconds},\ }\href {https://doi.org/10.1103/1qhb-r4fb} {\bibfield  {journal} {\bibinfo  {journal} {Phys. Rev. Res.}\ }\textbf {\bibinfo {volume} {7}},\ \bibinfo {pages} {043142} (\bibinfo {year} {2025})}\BibitemShut {NoStop}%
\bibitem [{\citenamefont {Swiadek}\ \emph {et~al.}(2024)Swiadek, F. and Shillito, R. and Magnard, P. and Remm, A. and Hellings, C. and Lacroix, N. and Ficheux, Q. and Zanuz, D. C. and Norris, G. J. and Blais, A. and Krinner, S. and Wallraff, A.}]{Swiadek2024}%
  \BibitemOpen
  \bibfield  {author} {\bibinfo {author} {Swiadek, F.}, \bibinfo {author} {Shillito, R.}, \bibinfo {author} {Magnard, P.}, \emph {et~al.},\ }\bibfield  {title} {\bibinfo {title} {Enhancing dispersive readout of superconducting qubits through dynamic control of the dispersive shift: Experiment and theory},\ }\href {https://doi.org/10.1103/PRXQuantum.5.040326} {\bibfield  {journal} {\bibinfo  {journal} {PRX Quantum}\ }\textbf {\bibinfo {volume} {5}},\ \bibinfo {pages} {040326} (\bibinfo {year} {2024})}\BibitemShut {NoStop}%
\bibitem [{\citenamefont {Magesan}\ \emph {et~al.}(2011)Magesan, Easwar and Gambetta, J. M. and Emerson, Joseph}]{Magesan2011}%
  \BibitemOpen
  \bibfield  {author} {\bibinfo {author} {Magesan, E.}, \bibinfo {author} {Gambetta, J.~M.},\ and\ \bibinfo {author} {Emerson, J.},\ }\bibfield  {title} {\bibinfo {title} {Scalable and robust randomized benchmarking of quantum processes},\ }\href {https://doi.org/10.1103/PhysRevLett.106.180504} {\bibfield  {journal} {\bibinfo  {journal} {Phys. Rev. Lett.}\ }\textbf {\bibinfo {volume} {106}},\ \bibinfo {pages} {180504} (\bibinfo {year} {2011})}\BibitemShut {NoStop}%
\bibitem [{\citenamefont {Epstein}\ \emph {et~al.}(2014)Jeffrey M. Epstein and Andrew W. Cross and Easwar Magesan and Jay M. Gambetta}]{Epstein2014}%
  \BibitemOpen
  \bibfield  {author} {\bibinfo {author} {Epstein, J.~M.}, \bibinfo {author} {Cross, A.~W.}, \bibinfo {author} {Magesan, E.},\ and\ \bibinfo {author} {Gambetta, J.~M.},\ }\bibfield  {title} {\bibinfo {title} {Investigating the limits of randomized benchmarking protocols},\ }\href {https://doi.org/10.1103/PhysRevA.89.062321} {\bibfield  {journal} {\bibinfo  {journal} {Phys. Rev. A}\ }\textbf {\bibinfo {volume} {89}},\ \bibinfo {pages} {062321} (\bibinfo {year} {2014})}\BibitemShut {NoStop}%
\bibitem [{\citenamefont {Magesan}\ \emph {et~al.}(2012)Magesan, Easwar and Gambetta, Jay M. and Johnson, B. R. and Ryan, Colm A. and Chow, Jerry M. and Merkel, Seth T. and da Silva, Marcus P. and Keefe, George A. and Rothwell, Mary B. and Ohki, Thomas A. and Ketchen, Mark B. and Steffen, M.}]{Magesan2012}%
  \BibitemOpen
  \bibfield  {author} {\bibinfo {author} {Magesan, E.}, \bibinfo {author} {Gambetta, J.~M.}, \bibinfo {author} {Johnson, B.~R.}, \emph {et~al.},\ }\bibfield  {title} {\bibinfo {title} {Efficient measurement of quantum gate error by interleaved randomized benchmarking},\ }\href {https://doi.org/10.1103/PhysRevLett.109.080505} {\bibfield  {journal} {\bibinfo  {journal} {Phys. Rev. Lett.}\ }\textbf {\bibinfo {volume} {109}},\ \bibinfo {pages} {080505} (\bibinfo {year} {2012})}\BibitemShut {NoStop}%
\bibitem [{\citenamefont {C\'orcoles}\ \emph {et~al.}(2013)C\'orcoles, A. D. and Gambetta, Jay M. and Chow, Jerry M. and Smolin, John A. and Ware, Matthew and Strand, Joel and Plourde, B. L. T. and Steffen, M.}]{Corcoles2013}%
  \BibitemOpen
  \bibfield  {author} {\bibinfo {author} {C\'orcoles, A.~D.}, \bibinfo {author} {Gambetta, J.~M.}, \bibinfo {author} {Chow, J.~M.}, \emph {et~al.},\ }\bibfield  {title} {\bibinfo {title} {Process verification of two-qubit quantum gates by randomized benchmarking},\ }\href {https://doi.org/10.1103/PhysRevA.87.030301} {\bibfield  {journal} {\bibinfo  {journal} {Phys. Rev. A}\ }\textbf {\bibinfo {volume} {87}},\ \bibinfo {pages} {030301} (\bibinfo {year} {2013})}\BibitemShut {NoStop}%
\bibitem [{\citenamefont {Barends}\ \emph {et~al.}(2014)Barends, R. and Kelly, J. and Megrant, A. and Veitia, A. and Sank, D. and Jeffrey, E. and White, T. C. and Mutus, J. and Fowler, A. G. and Campbell, B. and Chen, Y. and Chen, Z. and Chiaro, B. and Dunsworth, A. and Neill, C. and {O\'Malley}, P. and Roushan, P. and Vainsencher, A. and Wenner, J. and Korotkov, A. N. and Cleland, A. N. and Martinis, John M.}]{Barends2014}%
  \BibitemOpen
  \bibfield  {author} {\bibinfo {author} {Barends, R.}, \bibinfo {author} {Kelly, J.}, \bibinfo {author} {Megrant, A.}, \emph {et~al.},\ }\bibfield  {title} {\bibinfo {title} {Superconducting quantum circuits at the surface code threshold for fault tolerance},\ }\href {https://doi.org/10.1038/nature13171} {\bibfield  {journal} {\bibinfo  {journal} {Nature}\ }\textbf {\bibinfo {volume} {508}},\ \bibinfo {pages} {500} (\bibinfo {year} {2014})}\BibitemShut {NoStop}%
\bibitem [{\citenamefont {Krinner}\ \emph {et~al.}(2019)Krinner, S. and Storz, S. and Kurpiers, P. and Magnard, P. and Heinsoo, J. and Keller, R. and L{\"u}tolf, J. and Eichler, C. and Wallraff, A.}]{Krinner2019}%
  \BibitemOpen
  \bibfield  {author} {\bibinfo {author} {Krinner, S.}, \bibinfo {author} {Storz, S.}, \bibinfo {author} {Kurpiers, P.}, \emph {et~al.},\ }\bibfield  {title} {\bibinfo {title} {Engineering cryogenic setups for 100-qubit scale superconducting circuit systems},\ }\href {https://doi.org/10.1140/epjqt/s40507-019-0072-0} {\bibfield  {journal} {\bibinfo  {journal} {EPJ Quantum Technol.}\ }\textbf {\bibinfo {volume} {6}},\ \bibinfo {pages} {2} (\bibinfo {year} {2019})}\BibitemShut {NoStop}%
\bibitem [{\citenamefont {Macklin}\ \emph {et~al.}(2015)Macklin, C. and O'Brien, K. and Hover, D. and Schwartz, M. E. and Bolkhovsky, V. and Zhang, X. and Oliver, W. D. and Siddiqi, I.}]{Macklin2015}%
  \BibitemOpen
  \bibfield  {author} {\bibinfo {author} {Macklin, C.}, \bibinfo {author} {O'Brien, K.}, \bibinfo {author} {Hover, D.}, \emph {et~al.},\ }\bibfield  {title} {\bibinfo {title} {A near-quantum-limited {Josephson} traveling-wave parametric amplifier},\ }\href {https://doi.org/10.1126/science.aaa8525} {\bibfield  {journal} {\bibinfo  {journal} {Science}\ }\textbf {\bibinfo {volume} {350}},\ \bibinfo {pages} {307} (\bibinfo {year} {2015})}\BibitemShut {NoStop}%
\bibitem [{\citenamefont {Heinsoo}\ \emph {et~al.}(2018)Heinsoo, Johannes and Andersen, Christian Kraglund and Remm, Ants and Krinner, Sebastian and Walter, Theodore and Salath\'{e}, Yves and Gasparinetti, Simone and Besse, Jean-Claude and Poto\v{c}nik, Anton and Wallraff, Andreas and Eichler, Christopher}]{Heinsoo2018}%
  \BibitemOpen
  \bibfield  {author} {\bibinfo {author} {Heinsoo, J.}, \bibinfo {author} {Andersen, C.~K.}, \bibinfo {author} {Remm, A.}, \emph {et~al.},\ }\bibfield  {title} {\bibinfo {title} {Rapid high-fidelity multiplexed readout of superconducting qubits},\ }\href {https://doi.org/10.1103/PhysRevApplied.10.034040} {\bibfield  {journal} {\bibinfo  {journal} {Phys. Rev. Appl.}\ }\textbf {\bibinfo {volume} {10}},\ \bibinfo {pages} {034040} (\bibinfo {year} {2018})}\BibitemShut {NoStop}%
\bibitem [{\citenamefont {Bravyi}\ \emph {et~al.}(2021)Bravyi, Sergey and Sheldon, Sarah and Kandala, Abhinav and Mckay, David C. and Gambetta, Jay M.}]{Bravyi2021}%
  \BibitemOpen
  \bibfield  {author} {\bibinfo {author} {Bravyi, S.}, \bibinfo {author} {Sheldon, S.}, \bibinfo {author} {Kandala, A.}, \emph {et~al.},\ }\bibfield  {title} {\bibinfo {title} {Mitigating measurement errors in multiqubit experiments},\ }\href {https://doi.org/10.1103/PhysRevA.103.042605} {\bibfield  {journal} {\bibinfo  {journal} {Phys. Rev. A}\ }\textbf {\bibinfo {volume} {103}},\ \bibinfo {pages} {042605} (\bibinfo {year} {2021})}\BibitemShut {NoStop}%
\bibitem [{\citenamefont {Javadi-Abhari}\ \emph {et~al.}(2024)Javadi-Abhari, Ali and Treinish, Matthew and Krsulich, Kevin and Wood, Christopher J. and Lishman, Jake and Gacon, Julien and Martiel, Simon and Nation, Paul D. and Bishop, Lev S. and Cross, Andrew W. and Johnson, Blake R. and Gambetta, Jay M.}]{Javadi-Abhari2024}%
  \BibitemOpen
  \bibfield  {author} {\bibinfo {author} {Javadi-Abhari, A.}, \bibinfo {author} {Treinish, M.}, \bibinfo {author} {Krsulich, K.}, \emph {et~al.},\ }\bibfield  {title} {\bibinfo {title} {Quantum computing with qiskit},\ }\href {http://arxiv.org/abs/2405.08810} {\bibfield  {journal} {\bibinfo  {journal} {arXiv:2405.08810}\ } (\bibinfo {year} {2024})}\BibitemShut {NoStop}%
\bibitem [{\citenamefont {Nielsen}\ and\ \citenamefont {Chuang}(2010)Nielsen, Michael~A. and Chuang, Isaac~L.}]{Nielsen2010}%
  \BibitemOpen
  \bibfield  {author} {\bibinfo {author} {Nielsen, M.~A.}\ and\ \bibinfo {author} {Chuang, I.~L.},\ }\href@noop {} {\emph {\bibinfo {title} {Quantum Computation and Quantum Information}}},\ \bibinfo {edition} {{10th anniversary}}\ ed.\ (\bibinfo  {publisher} {Cambridge University Press},\ \bibinfo {address} {New York, USA},\ \bibinfo {year} {2010})\BibitemShut {NoStop}%
\bibitem [{\citenamefont {Chen}\ \emph {et~al.}(2021)Chen, Zijun and Satzinger, Kevin J. and Atalaya, Juan and Korotkov, Alexander N. and Dunsworth, Andrew and Sank, Daniel and Quintana, Chris and McEwen, Matt and Barends, Rami and Klimov, Paul V. and Hong, Sabrina and Jones, Cody and Petukhov, Andre and Kafri, Dvir and Demura, Sean and Burkett, Brian and Gidney, Craig and Fowler, Austin G. and Paler, Alexandru and Putterman, Harald and Aleiner, Igor and Arute, Frank and Arya, Kunal and Babbush, Ryan and Bardin, Joseph C. and Bengtsson, Andreas and Bourassa, Alexandre and Broughton, Michael and Buckley, Bob B. and Buell, David A. and Bushnell, Nicholas and Chiaro, Benjamin and Collins, Roberto and Courtney, William and Derk, Alan R. and Eppens, Daniel and Erickson, Catherine and Farhi, Edward and Foxen, Brooks and Giustina, Marissa and Greene, Ami and Gross, Jonathan A. and Harrigan, Matthew P. and Harrington, Sean D. and Hilton, Jeremy and Ho, Alan and Huang, Trent and Huggins, William J. and Ioffe, L. B.
  and Isakov, Sergei V. and Jeffrey, Evan and Jiang, Zhang and Kechedzhi, Kostyantyn and Kim, Seon and Kitaev, Alexei and Kostritsa, Fedor and Landhuis, David and Laptev, Pavel and Lucero, Erik and Martin, Orion and McClean, Jarrod R. and McCourt, Trevor and Mi, Xiao and Miao, Kevin C. and Mohseni, Masoud and Montazeri, Shirin and Mruczkiewicz, Wojciech and Mutus, Josh and Naaman, Ofer and Neeley, Matthew and Neill, Charles and Newman, Michael and Niu, Murphy Yuezhen and O’Brien, Thomas E. and Opremcak, Alex and Ostby, Eric and Pató, Bálint and Redd, Nicholas and Roushan, Pedram and Rubin, Nicholas C. and Shvarts, Vladimir and Strain, Doug and Szalay, Marco and Trevithick, Matthew D. and Villalonga, Benjamin and White, Theodore and Yao, Z. Jamie and Yeh, Ping and Yoo, Juhwan and Zalcman, Adam and Neven, Hartmut and Boixo, Sergio and Smelyanskiy, Vadim and Chen, Yu and Megrant, Anthony and Kelly, Julian}]{Chen2021p}%
  \BibitemOpen
  \bibfield  {author} {\bibinfo {author} {Chen, Z.}, \bibinfo {author} {Satzinger, K.~J.}, \bibinfo {author} {Atalaya, J.}, \emph {et~al.},\ }\bibfield  {title} {\bibinfo {title} {Exponential suppression of bit or phase errors with cyclic error correction},\ }\href {https://doi.org/10.1038/s41586-021-03588-y} {\bibfield  {journal} {\bibinfo  {journal} {Nature}\ }\textbf {\bibinfo {volume} {595}},\ \bibinfo {pages} {383} (\bibinfo {year} {2021})}\BibitemShut {NoStop}%
\bibitem [{\citenamefont {McKay}\ \emph {et~al.}(2017)McKay, David C. and Wood, Christopher J. and Sheldon, Sarah and Chow, Jerry M. and Gambetta, Jay M.}]{McKay2017}%
  \BibitemOpen
  \bibfield  {author} {\bibinfo {author} {McKay, D.~C.}, \bibinfo {author} {Wood, C.~J.}, \bibinfo {author} {Sheldon, S.}, \emph {et~al.},\ }\bibfield  {title} {\bibinfo {title} {Efficient {$Z$} gates for quantum computing},\ }\href {https://doi.org/10.1103/PhysRevA.96.022330} {\bibfield  {journal} {\bibinfo  {journal} {Phys. Rev. A}\ }\textbf {\bibinfo {volume} {96}},\ \bibinfo {pages} {022330} (\bibinfo {year} {2017})}\BibitemShut {NoStop}%
\bibitem [{\citenamefont {Rist\`e}\ \emph {et~al.}(2012)Rist\`e, D. and van Leeuwen, J. G. and Ku, H.-S. and Lehnert, K. W. and DiCarlo, L.}]{Riste2012}%
  \BibitemOpen
  \bibfield  {author} {\bibinfo {author} {Rist\`e, D.}, \bibinfo {author} {van Leeuwen, J.~G.}, \bibinfo {author} {Ku, H.-S.}, \emph {et~al.},\ }\bibfield  {title} {\bibinfo {title} {Initialization by measurement of a superconducting quantum bit circuit},\ }\href {https://doi.org/10.1103/PhysRevLett.109.050507} {\bibfield  {journal} {\bibinfo  {journal} {Phys. Rev. Lett.}\ }\textbf {\bibinfo {volume} {109}},\ \bibinfo {pages} {050507} (\bibinfo {year} {2012})}\BibitemShut {NoStop}%
\bibitem [{\citenamefont {Johnson}\ \emph {et~al.}(2012)Johnson, J. E. and Macklin, C. and Slichter, D. H. and Vijay, R. and Weingarten, E. B. and Clarke, John and Siddiqi, I.}]{Johnson2012}%
  \BibitemOpen
  \bibfield  {author} {\bibinfo {author} {Johnson, J.~E.}, \bibinfo {author} {Macklin, C.}, \bibinfo {author} {Slichter, D.~H.}, \emph {et~al.},\ }\bibfield  {title} {\bibinfo {title} {Heralded state preparation in a superconducting qubit},\ }\href {https://doi.org/10.1103/PhysRevLett.109.050506} {\bibfield  {journal} {\bibinfo  {journal} {Phys. Rev. Lett.}\ }\textbf {\bibinfo {volume} {109}},\ \bibinfo {pages} {050506} (\bibinfo {year} {2012})}\BibitemShut {NoStop}%
\bibitem [{\citenamefont {Carr}\ and\ \citenamefont {Purcell}(1954)Carr, H. Y. and Purcell, E. M.}]{Carr1954}%
  \BibitemOpen
  \bibfield  {author} {\bibinfo {author} {Carr, H.~Y.}\ and\ \bibinfo {author} {Purcell, E.~M.},\ }\bibfield  {title} {\bibinfo {title} {Effects of diffusion on free precession in nuclear magnetic resonance experiments},\ }\href {https://doi.org/10.1103/PhysRev.94.630} {\bibfield  {journal} {\bibinfo  {journal} {Phys. Rev.}\ }\textbf {\bibinfo {volume} {94}},\ \bibinfo {pages} {630} (\bibinfo {year} {1954})}\BibitemShut {NoStop}%
\bibitem [{\citenamefont {Meiboom}\ and\ \citenamefont {Gill}(1958)Meiboom, S. and Gill, D.}]{Meiboom1958}%
  \BibitemOpen
  \bibfield  {author} {\bibinfo {author} {Meiboom, S.}\ and\ \bibinfo {author} {Gill, D.},\ }\bibfield  {title} {\bibinfo {title} {Modified spin-echo method for measuring nuclear relaxation times},\ }\href {https://doi.org/10.1063/1.1716296} {\bibfield  {journal} {\bibinfo  {journal} {Rev. Sci. Instrum.}\ }\textbf {\bibinfo {volume} {29}},\ \bibinfo {pages} {688} (\bibinfo {year} {1958})}\BibitemShut {NoStop}%
\bibitem [{\citenamefont {Somoroff}\ \emph {et~al.}(2023)Somoroff, Aaron and Ficheux, Quentin and Mencia, Raymond A. and Xiong, Haonan and Kuzmin, Roman and Manucharyan, Vladimir E.}]{Somoroff2023b}%
  \BibitemOpen
  \bibfield  {author} {\bibinfo {author} {Somoroff, A.}, \bibinfo {author} {Ficheux, Q.}, \bibinfo {author} {Mencia, R.~A.}, \emph {et~al.},\ }\bibfield  {title} {\bibinfo {title} {Millisecond coherence in a superconducting qubit},\ }\href {https://doi.org/10.1103/PhysRevLett.130.267001} {\bibfield  {journal} {\bibinfo  {journal} {Phys. Rev. Lett.}\ }\textbf {\bibinfo {volume} {130}},\ \bibinfo {pages} {267001} (\bibinfo {year} {2023})}\BibitemShut {NoStop}%
\bibitem [{\citenamefont {Li}\ \emph {et~al.}(2023)Li, Zhiyuan and Liu, Pei and Zhao, Peng and Mi, Zhenyu and Xu, Huikai and Liang, Xuehui and Su, Tang and Sun, Weijie and Xue, Guangming and Zhang, Jing-Ning and Liu, Weiyang and Jin, Yirong and Yu, Haifeng}]{Li2023n}%
  \BibitemOpen
  \bibfield  {author} {\bibinfo {author} {Li, Z.}, \bibinfo {author} {Liu, P.}, \bibinfo {author} {Zhao, P.}, \emph {et~al.},\ }\bibfield  {title} {\bibinfo {title} {Error per single-qubit gate below 10-4 in a superconducting qubit},\ }\href {https://doi.org/10.1038/s41534-023-00781-x} {\bibfield  {journal} {\bibinfo  {journal} {npj Quantum Inf.}\ }\textbf {\bibinfo {volume} {9}},\ \bibinfo {pages} {111} (\bibinfo {year} {2023})}\BibitemShut {NoStop}%
\bibitem [{\citenamefont {Rower}\ \emph {et~al.}(2024)Rower, David A. and Ding, Leon and Zhang, Helin and Hays, Max and An, Junyoung and Harrington, Patrick M. and Rosen, Ilan T. and Gertler, Jeffrey M. and Hazard, Thomas M. and Niedzielski, Bethany M. and Schwartz, Mollie E. and Gustavsson, Simon and Serniak, Kyle and Grover, Jeffrey A. and Oliver, William D.}]{Rower2024}%
  \BibitemOpen
  \bibfield  {author} {\bibinfo {author} {Rower, D.~A.}, \bibinfo {author} {Ding, L.}, \bibinfo {author} {Zhang, H.}, \emph {et~al.},\ }\bibfield  {title} {\bibinfo {title} {Suppressing counter-rotating errors for fast single-qubit gates with fluxonium},\ }\href {https://doi.org/10.1103/PRXQuantum.5.040342} {\bibfield  {journal} {\bibinfo  {journal} {PRX Quantum}\ }\textbf {\bibinfo {volume} {5}},\ \bibinfo {pages} {040342} (\bibinfo {year} {2024})}\BibitemShut {NoStop}%
\bibitem [{\citenamefont {Marxer}\ \emph {et~al.}(2025)Fabian Marxer and Jakub Mrożek and Joona Andersson and Leonid Abdurakhimov and Janos Adam and Ville Bergholm and Rohit Beriwal and Chun Fai Chan and Saga Dahl and Soumya Ranjan Das and Frank Deppe and Olexiy Fedorets and Zheming Gao and Alejandro Gomez Frieiro and Daria Gusenkova and Andrew Guthrie and Tuukka Hiltunen and Hao Hsu and Eric Hyyppä and Joni Ikonen and Sinan Inel and Shan W. Jolin and Azad Karis and Seung-Goo Kim and William Kindel and Anton Komlev and Miikka Koistinen and Roope Kokkoniemi and Snigdha Kumar and Hsiang-Sheng Ku and Julia Lamprich and Sami Laine and Alessandro Landra and Lan-Hsuan Lee and Nizar Lethif and Per Liebermann and Wei Liu and Kunal Mitra and Tuomas Mylläri and Caspar Ockeloen-Korppi and Tuure Orell and Alexander Plyshch and Jukka Räbinä and Arthur Rebello and Michael Renger and Outi Reentilä and Jussi Ritvas and Sampo Saarinen and Otto Salmenkivi and Matthew Sarsby and Mykhailo Savytskyi and Ville Selinmaa and
  Matthew Steggles and Eelis Takala and Ivan Takmakov and Brian Tarasinski and Jani Tuorila and Alpo Välimaa and Jeroen Verjauw and Jaap Wesdorp and Nicola Wurz and Wei Qiu and Lihuang Zhu and Juha Hassel and Johannes Heinsoo and Attila Geresdi and Antti Vepsäläinen}]{Marxer2025}%
  \BibitemOpen
  \bibfield  {author} {\bibinfo {author} {Marxer, F.}, \bibinfo {author} {Mrożek, J.}, \bibinfo {author} {Andersson, J.}, \emph {et~al.},\ }\bibfield  {title} {\bibinfo {title} {{Above 99.9\% Fidelity Single-Qubit Gates, Two-Qubit Gates, and Readout in a Single Superconducting Quantum Device}},\ }\href {https://arxiv.org/abs/2508.16437} {\bibfield  {journal} {\bibinfo  {journal} {arXiv:2508.16437}\ } (\bibinfo {year} {2025})}\BibitemShut {NoStop}%
\bibitem [{\citenamefont {Ding}\ \emph {et~al.}(2023)Leon Ding and Max Hays and Youngkyu Sung and Bharath Kannan and Junyoung An and Agustin Di Paolo and Amir H. Karamlou and Thomas M. Hazard and Kate Azar and David K. Kim and Bethany M. Niedzielski and Alexander Melville and Mollie E. Schwartz and Jonilyn L. Yoder and Terry P. Orlando and Simon Gustavsson and Jeffrey A. Grover and Kyle Serniak and William D. Oliver}]{Ding2023}%
  \BibitemOpen
  \bibfield  {author} {\bibinfo {author} {Ding, L.}, \bibinfo {author} {Hays, M.}, \bibinfo {author} {Sung, Y.}, \emph {et~al.},\ }\bibfield  {title} {\bibinfo {title} {High-fidelity, frequency-flexible two-qubit fluxonium gates with a transmon coupler},\ }\href {https://doi.org/10.1103/physrevx.13.031035} {\bibfield  {journal} {\bibinfo  {journal} {Phys. Rev. X}\ }\textbf {\bibinfo {volume} {13}},\ \bibinfo {pages} {031035} (\bibinfo {year} {2023})}\BibitemShut {NoStop}%
\bibitem [{\citenamefont {Lin}\ \emph {et~al.}(2025)Lin, Wei-Ju and Cho, Hyunheung and Chen, Yinqi and Vavilov, Maxim G. and Wang, Chen and Manucharyan, Vladimir E.}]{Lin2024a}%
  \BibitemOpen
  \bibfield  {author} {\bibinfo {author} {Lin, W.-J.}, \bibinfo {author} {Cho, H.}, \bibinfo {author} {Chen, Y.}, \emph {et~al.},\ }\bibfield  {title} {\bibinfo {title} {{24 Days-Stable CNOT Gate on Fluxonium Qubits with Over 99.9\% Fidelity}},\ }\href {https://doi.org/10.1103/PRXQuantum.6.010349} {\bibfield  {journal} {\bibinfo  {journal} {PRX Quantum}\ }\textbf {\bibinfo {volume} {6}},\ \bibinfo {pages} {010349} (\bibinfo {year} {2025})}\BibitemShut {NoStop}%
\bibitem [{\citenamefont {Li}\ \emph {et~al.}(2024)Li, Rui and Kubo, Kentaro and Ho, Yinghao and Yan, Zhiguang and Nakamura, Yasunobu and Goto, Hayato}]{Li2024c}%
  \BibitemOpen
  \bibfield  {author} {\bibinfo {author} {Li, R.}, \bibinfo {author} {Kubo, K.}, \bibinfo {author} {Ho, Y.}, \emph {et~al.},\ }\bibfield  {title} {\bibinfo {title} {Realization of high-fidelity cz gate based on a double-transmon coupler},\ }\href {https://doi.org/10.1103/PhysRevX.14.041050} {\bibfield  {journal} {\bibinfo  {journal} {Phys. Rev. X}\ }\textbf {\bibinfo {volume} {14}},\ \bibinfo {pages} {041050} (\bibinfo {year} {2024})}\BibitemShut {NoStop}%
\bibitem [{\citenamefont {{IBM Quantum}}(2025){IBM Quantum}}]{IBM2025}%
  \BibitemOpen
  \bibfield  {author} {\bibinfo {author} {{IBM Quantum}},\ }\href {https://quantum.cloud.ibm.com/docs/en/guides/processor-types} {\bibinfo {title} {https://quantum.ibm.com/}} (\bibinfo {year} {2025})\BibitemShut {NoStop}%
\bibitem [{\citenamefont {Spring}\ \emph {et~al.}(2025)Spring, Peter A. and Milanovic, Luka and Sunada, Yoshiki and Wang, Shiyu and van Loo, Arjan F. and Tamate, Shuhei and Nakamura, Yasunobu}]{Spring2025}%
  \BibitemOpen
  \bibfield  {author} {\bibinfo {author} {Spring, P.~A.}, \bibinfo {author} {Milanovic, L.}, \bibinfo {author} {Sunada, Y.}, \emph {et~al.},\ }\bibfield  {title} {\bibinfo {title} {Fast multiplexed superconducting-qubit readout with intrinsic {Purcell} filtering using a multiconductor transmission line},\ }\href {https://doi.org/10.1103/PRXQuantum.6.020345} {\bibfield  {journal} {\bibinfo  {journal} {PRX Quantum}\ }\textbf {\bibinfo {volume} {6}},\ \bibinfo {pages} {020345} (\bibinfo {year} {2025})}\BibitemShut {NoStop}%
\bibitem [{\citenamefont {Wang}\ \emph {et~al.}(2025{\natexlab{a}})Wang, Can and Liu, Feng-Ming and Chen, He and Du, Yi-Fei and Ying, Chong and Wang, Jian-Wen and Huo, Yong-Heng and Peng, Cheng-Zhi and Zhu, Xiaobo and Chen, Ming-Cheng and Lu, Chao-Yang and Pan, Jian-Wei}]{Wang2025r}%
  \BibitemOpen
  \bibfield  {author} {\bibinfo {author} {Wang, C.}, \bibinfo {author} {Liu, F.-M.}, \bibinfo {author} {Chen, H.}, \emph {et~al.},\ }\bibfield  {title} {\bibinfo {title} {Longitudinal and nonlinear coupling for high-fidelity readout of a superconducting qubit},\ }\href {https://doi.org/10.1103/98n9-13y4} {\bibfield  {journal} {\bibinfo  {journal} {Phys. Rev. Lett.}\ }\textbf {\bibinfo {volume} {135}},\ \bibinfo {pages} {060803} (\bibinfo {year} {2025}{\natexlab{a}})}\BibitemShut {NoStop}%
\bibitem [{\citenamefont {Place}\ \emph {et~al.}(2021)Place, A. P. M. and Rodgers, L. V. H. and Mundada, P. and Smitham, B. M. and Fitzpatrick, M. and Leng, Z. and Premkumar, A. and Bryon, J. and Sussman, S. and Cheng, G. and Madhavan, T. and Babla, H. K. and Jaeck, B. and Gyenis, A. and Yao, N. and Cava, R. J. and de Leon, N. P. and Houck, A. A.}]{Place2021}%
  \BibitemOpen
  \bibfield  {author} {\bibinfo {author} {Place, A. P.~M.}, \bibinfo {author} {Rodgers, L. V.~H.}, \bibinfo {author} {Mundada, P.}, \emph {et~al.},\ }\bibfield  {title} {\bibinfo {title} {New material platform for superconducting transmon qubits with coherence times exceeding 0.3 milliseconds},\ }\href {https://doi.org/10.1038/s41467-021-22030-5} {\bibfield  {journal} {\bibinfo  {journal} {Nat. Commun.}\ }\textbf {\bibinfo {volume} {12}},\ \bibinfo {pages} {1779} (\bibinfo {year} {2021})}\BibitemShut {NoStop}%
\bibitem [{\citenamefont {Wang}\ \emph {et~al.}(2022)Wang, Chenlu and Li, Xuegang and Xu, Huikai and Li, Zhiyuan and Wang, Junhua and Yang, Zhen and Mi, Zhenyu and Liang, Xuehui and Su, Tang and Yang, Chuhong and Wang, Guangyue and Wang, Wenyan and Li, Yongchao and Chen, Mo and Li, Chengyao and Linghu, Kehuan and Han, Jiaxiu and Zhang, Yingshan and Feng, Yulong and Song, Yu and Ma, Teng and Zhang, Jingning and Wang, Ruixia and Zhao, Peng and Liu, Weiyang and Xue, Guangming and Jin, Yirong and Yu, Haifeng}]{Wang2022a}%
  \BibitemOpen
  \bibfield  {author} {\bibinfo {author} {Wang, C.}, \bibinfo {author} {Li, X.}, \bibinfo {author} {Xu, H.}, \emph {et~al.},\ }\bibfield  {title} {\bibinfo {title} {Towards practical quantum computers: transmon qubit with a lifetime approaching 0.5 milliseconds},\ }\href {https://doi.org/10.1038/s41534-021-00510-2} {\bibfield  {journal} {\bibinfo  {journal} {npj Quantum Inf.}\ }\textbf {\bibinfo {volume} {8}},\ \bibinfo {pages} {3} (\bibinfo {year} {2022})}\BibitemShut {NoStop}%
\bibitem [{\citenamefont {Tuokkola}\ \emph {et~al.}(2025)Tuokkola, Mikko and Sunada, Yoshiki and Kivij{\"a}rvi, Heidi and Albanese, Jonatan and Gr{\"o}nberg, Leif and Kaikkonen, Jukka-Pekka and Vesterinen, Visa and Govenius, Joonas and M{\"o}tt{\"o}nen, Mikko}]{Tuokkola2025}%
  \BibitemOpen
  \bibfield  {author} {\bibinfo {author} {Tuokkola, M.}, \bibinfo {author} {Sunada, Y.}, \bibinfo {author} {Kivij{\"a}rvi, H.}, \emph {et~al.},\ }\bibfield  {title} {\bibinfo {title} {Methods to achieve near-millisecond energy relaxation and dephasing times for a superconducting transmon qubit},\ }\href {https://doi.org/https://doi.org/10.1038/s41467-025-61126-0} {\bibfield  {journal} {\bibinfo  {journal} {Nat. Commun.}\ }\textbf {\bibinfo {volume} {16}},\ \bibinfo {pages} {5421} (\bibinfo {year} {2025})}\BibitemShut {NoStop}%
\bibitem [{\citenamefont {Wang}\ \emph {et~al.}(2025{\natexlab{b}})Wang, Fei and Lu, Kannan and Zhan, Huijuan and Ma, Lu and Wu, Feng and Sun, Hantao and Deng, Hao and Bai, Yang and Bao, Feng and Chang, Xu and Gao, Ran and Gao, Xun and Gong, Guicheng and Hu, Lijuan and Hu, Ruizi and Ji, Honghong and Ma, Xizheng and Mao, Liyong and Song, Zhijun and Tang, Chengchun and Wang, Hongcheng and Wang, Tenghui and Wang, Ziang and Xia, Tian and Xu, Hongxin and Zhan, Ze and Zhang, Gengyan and Zhou, Tao and Zhu, Mengyu and Zhu, Qingbin and Zhu, Shasha and Zhu, Xing and Shi, Yaoyun and Zhao, Hui-Hai and Deng, Chunqing}]{Wang2025g}%
  \BibitemOpen
  \bibfield  {author} {\bibinfo {author} {Wang, F.}, \bibinfo {author} {Lu, K.}, \bibinfo {author} {Zhan, H.}, \emph {et~al.},\ }\bibfield  {title} {\bibinfo {title} {High-coherence fluxonium qubits manufactured with a wafer-scale-uniformity process},\ }\href {https://doi.org/10.1103/PhysRevApplied.23.044064} {\bibfield  {journal} {\bibinfo  {journal} {Phys. Rev. Appl.}\ }\textbf {\bibinfo {volume} {23}},\ \bibinfo {pages} {044064} (\bibinfo {year} {2025}{\natexlab{b}})}\BibitemShut {NoStop}%
\bibitem [{\citenamefont {Bland}\ \emph {et~al.}(2025)Bland, Matthew P. and Bahrami, Faranak and Martinez, Jeronimo G. C. and Prestegaard, Paal H. and Smitham, Basil M. and Joshi, Atharv and Hedrick, Elizabeth and Kumar, Shashwat and Yang, Ambrose and Pakpour-Tabrizi, Alexander C. and Jindal, Apoorv and Chang, Ray D. and Cheng, Guangming and Yao, Nan and Cava, Robert J. and de Leon, Nathalie P. and Houck, Andrew A.}]{Bland2025}%
  \BibitemOpen
  \bibfield  {author} {\bibinfo {author} {Bland, M.~P.}, \bibinfo {author} {Bahrami, F.}, \bibinfo {author} {Martinez, J. G.~C.}, \emph {et~al.},\ }\bibfield  {title} {\bibinfo {title} {Millisecond lifetimes and coherence times in 2d transmon qubits},\ }\href {https://doi.org/10.1038/s41586-025-09687-4} {\bibfield  {journal} {\bibinfo  {journal} {Nature}\ }\textbf {\bibinfo {volume} {647}},\ \bibinfo {pages} {343} (\bibinfo {year} {2025})}\BibitemShut {NoStop}%
\bibitem [{\citenamefont {Poshtvan}\ \emph {et~al.}(2025)Poshtvan, Abbas and Lapiha, Oleksandra and Doosti, Mina and Leichtle, Dominik and Music, Luka and Kashefi, Elham}]{Poshtvan2025}%
  \BibitemOpen
  \bibfield  {author} {\bibinfo {author} {Poshtvan, A.}, \bibinfo {author} {Lapiha, O.}, \bibinfo {author} {Doosti, M.}, \emph {et~al.},\ }\bibfield  {title} {\bibinfo {title} {Selectively blind quantum computation},\ }\href {https://arxiv.org/abs/2504.17612} {\bibfield  {journal} {\bibinfo  {journal} {arXiv:2504.17612}\ } (\bibinfo {year} {2025})}\BibitemShut {NoStop}%
\bibitem [{\citenamefont {Baranes}\ \emph {et~al.}(2025)Baranes, Gefen and Wang, Iria W and Machado, Francisco and Suleymanzade, Aziza and Stas, Pieter-Jan and Wei, Yan-Cheng and Yelin, Susanne F and Borregaard, Johannes and Lukin, Mikhail D}]{Baranes2025a}%
  \BibitemOpen
  \bibfield  {author} {\bibinfo {author} {Baranes, G.}, \bibinfo {author} {Wang, I.~W.}, \bibinfo {author} {Machado, F.}, \emph {et~al.},\ }\bibfield  {title} {\bibinfo {title} {{Designing Fault-Tolerant Blind Quantum Computation}},\ }\href {https://arxiv.org/abs/2505.21621} {\bibfield  {journal} {\bibinfo  {journal} {arXiv:2505.21621}\ } (\bibinfo {year} {2025})}\BibitemShut {NoStop}%
\bibitem [{\citenamefont {Raussendorf}\ \emph {et~al.}(2006)R. Raussendorf and J. Harrington and K. Goyal}]{Raussendorf2006}%
  \BibitemOpen
  \bibfield  {author} {\bibinfo {author} {Raussendorf, R.}, \bibinfo {author} {Harrington, J.},\ and\ \bibinfo {author} {Goyal, K.},\ }\bibfield  {title} {\bibinfo {title} {A fault-tolerant one-way quantum computer},\ }\href {https://doi.org/https://doi.org/10.1016/j.aop.2006.01.012} {\bibfield  {journal} {\bibinfo  {journal} {Annals of Physics}\ }\textbf {\bibinfo {volume} {321}},\ \bibinfo {pages} {2242} (\bibinfo {year} {2006})}\BibitemShut {NoStop}%
\end{thebibliography}

\end{document}